\documentclass[traditabstract]{aa}  
\usepackage{graphicx}
\usepackage{txfonts}

\begin{document}

\title{Hot subdwarf stars in close-up view}

\subtitle{I. Rotational properties of subdwarf B stars in close binary systems 
and nature of their unseen companions
\thanks{Based on observations at the Paranal Observatory of the European 
Southern Observatory for programmes number 165.H-0588(A), 167.D-0407(A), 
068.D-0483(A), 069.D-0534(A), 070.D-0334(A), 071.D-0380(A), 071.D-0383(A) and 
382.D-0841(A). Based on observations at the La Silla Observatory of the 
European Southern Observatory for programmes number 073.D-0495(A), 
074.B-0455(A) and 077.D-0515(A). Some of the data used in this work were 
obtained at the Hobby-Eberly Telescope (HET), which is a joint project of the
 University of Texas at Austin, the Pennsylvania State University, Stanford 
 University, Ludwig-Maximilians-Universit\"at M\"unchen, and 
 Georg-August-Universit\"at G\"ottingen, for programmes number UT07-2-004 and
  UT07-3-005. The HET is named in honor of its principal benefactors, 
  William P. Hobby and Robert E. Eberly. Based on observations collected at 
  the Centro Astron\'omico Hispano Alem\'an (CAHA) at Calar Alto, operated 
  jointly by the Max-Planck Institut f\"ur Astronomie and the Instituto de
   Astrof\'isica de Andaluc\'ia (CSIC). Some of the data presented here were 
   obtained at the W.M. Keck Observatory, which is operated as a scientific 
   partnership among the California Institute of Technology, the University of 
   California, and the National Aeronautics and Space Administration. The 
   Observatory was made possible by the generous financial support of the 
   W.M. Keck Foundation. Some of the data used in this work were obtained at the Palomar Observatory, owned and operated by the California Institute of Technology. Based on observations with the William Herschel Telescope operated by the Isaac Newton Group at the Observatorio del Roque de los Muchachos of the Instituto de Astrofisica de Canarias on the island of La Palma, Spain.}
}

\author{S. Geier \inst{1}
   \and U. Heber \inst{1}
   \and Ph. Podsiadlowski \inst{2}
   \and H. Edelmann \inst{1}
   \and R. Napiwotzki \inst{3}
   \and T. Kupfer \inst{1}
   \and S. M\"uller \inst{1}
   }

\offprints{S.\,Geier,\\ \email{geier@sternwarte.uni-erlangen.de}}

\institute{Dr. Karl Remeis-Observatory \& ECAP, Astronomical Institute,
Friedrich-Alexander University Erlangen-Nuremberg, Sternwartstr. 7, D 96049 Bamberg, Germany 
   \and Department of Astrophysics, University of Oxford, Keble Road, Oxford 
   OX1 3RH, UK
   \and Centre of Astrophysics Research, University of Hertfordshire, College 
   Lane, Hatfield AL10 9AB, UK}

\date{Received \ Accepted}

\abstract{The origin of hot subdwarf B stars (sdBs) is still unclear. About 
half of the known sdBs are in close binary systems for which 
common envelope ejection is the most likely formation channel. Little is 
known about this dynamic phase of binary evolution. Since most of the known 
sdB systems are single-lined spectroscopic binaries, it is difficult to derive
masses and unravel the companions' nature, which is the aim of this paper.\\
Due to the tidal influence of the companion in close binary 
systems, the rotation of the primary becomes synchronised to its orbital 
motion. In this case it is possible to constrain the mass of the companion, if the 
primary mass, its projected rotational velocity as well as its surface gravity 
are known. For the first time we measured the projected rotational velocities 
of a large sdB binary sample from high resolution spectra. We analysed a sample of 51 sdB stars in close binaries, 40 of which have  
known orbital parameters comprising half of all such systems known today.\\
Synchronisation in sdB binaries is discussed both from the 
theoretical and the observational point of view. The masses and the nature of the unseen companions could be constrained in 31 
cases. We found orbital synchronisation most likely to be established in binaries with orbital 
periods shorter than $1.2\,{\rm d}$. Only in five cases it was impossible to
decide whether the sdB's companion is a white dwarf or an M dwarf. The 
companions to seven sdBs could be clearly identified as late M stars. One 
binary may have a brown dwarf companion. The unseen companions of nine sdBs are white dwarfs with typical masses. The mass of one white dwarf companion is very low. In eight cases (including the well known system KPD1930$+$2752) 
the companion mass exceeds $0.9\,M_{\rm \odot}$, four of which even
exceed the Chandrasekhar limit indicating that they 
may be neutron stars. Even stellar mass black holes are possible for the most massive
companions. The distribution of the inclinations of the systems with low mass companions appears to 
be consistent with expectations, whereas a lack of high inclinations becomes 
obvious for the massive systems.
We show that the formation of such systems can be explained with common envelope evolution and present an appropriate formation channel including two phases of unstable mass transfer and one supernova explosion. 
 The sample also contains a candidate post-RGB star, which rotates fast despite its long orbital period. The post-RGB stars are expected to spin-up caused by their ongoing contraction. The age of the sdB is another important factor. If the EHB star is too young, the synchronisation process might not be finished yet. Estimating the ages of the target stars from their positions on the EHB band, we found PG\,2345$+$318, which is known not to be synchronised, to lie near the zero-age extreme horizontal branch as are the massive candidates PG\,1232$-$136, PG\,1432$+$159 and PG\,1101$+$249. These star may possibly be too young to have reached synchronisation.\\
The derived large fraction of putative massive sdB binary systems in low inclination orbits 
is inconsistent with theoretical predictions. Even if we dismiss three candidates because they may be too young and assume
that the other sdB primaries are of low mass, PG\,1743$+$477 and, in particular, HE\,0532$-$4503 remain as candidates whose companions may have masses close to or above the Chandrasekhar limit. X-ray observations and accurate photometry
are suggested to clarify their nature. As high inclination systems must also exist, an appropriate survey has already been launched to find such binaries.

}

\maketitle

\section{Introduction \label{sec:intro}}

Subluminous B stars (sdBs) show 
similar colours and spectral characteristics as main sequence stars of 
spectral type B, but are much less luminous. 
Compared to main sequence B stars the hydrogen Balmer lines in the spectra 
of sdBs are stronger while the helium lines are much weaker (if present at
all) 
for the colour. The strong line broadening and the early confluence of the
Balmer series is caused by the  
high surface gravities ($\log\,g\simeq5.0-6.0$) of these compact stars 
($R_{\rm sdB}\simeq0.1-0.3\,R_{\rm \odot}$). 
Subluminous B stars are considered to be helium core burning stars with 
very thin hydrogen envelopes and masses of about half a solar mass (Heber \cite{heber1})
located at the extreme end of the horizontal branch (EHB). 

Subdwarf B stars are 
found in all Galactic stellar populations and are sufficiently common to 
account
for the UV-upturn of early-type galaxies.
Understanding the origin of the UV-upturn phenomenon hence has to await a
proper understanding of  
the origin of the sdB stars themselves. 

The discovery of short-period multi-periodic pulsations 
in some sdBs provided an excellent opportunity to probe the
interiors of these stars using the tools of asteroseismology. 
They were theoretically predicted by
Charpinet et al. (\cite{charpinet96}) at around the same time as they were 
observed by
Kilkenny et al. (\cite{kilkenny}). 
They are
characterised by low-amplitude, multi-periodic, short-period ($80-600\,{\rm s}$)
light variations that are due to pressure ($p$)-mode
oscillations. 
A second family of pulsating sdB stars was discovered by Green et al. (\cite{green2}),
again showing low-amplitude, multi-periodic pulsations, but periods are longer ($2000-9000\,{\rm s}$) and are identified as 
gravity ($g$) modes. An important recent achievement of sdB asteroseismology 
is the determination of the most
fundamental parameter of a star, 
i.e. its mass (for a review see Fontaine et al. \cite{fontaine}).

The origin of EHB stars, however, is wrapped in mystery (see Heber
\cite{heber09} for a review). 
The problem is how some kind of mass loss mechanism in the
progenitor manages
to remove all but a tiny fraction of the hydrogen envelope at about
 the same time as 
the helium core has attained the mass ($\sim0.5$
M$_\odot$) required for the helium flash.
This requires enhanced mass loss, e.g. due to helium mixing driven by
internal rotation (Sweigart \cite{sweigart97}) or at the helium flash itself.

Mengel et al. (\cite{mengel76}) demonstrated that the required strong mass loss
can occur in a close-binary system. The progenitor of the sdB star has
to fill its Roche lobe near the tip of the red-giant branch (RGB) 
to lose most of its hydrogen-rich envelope. The merger of binary white dwarfs was investigated 
by Webbink (\cite{webbink}) who showed that an EHB star can form when two helium core
white dwarfs merge and the product is sufficiently massive to ignite helium.

Interest in the binary scenario was revived, 
when Maxted et al. (\cite{maxted2}) determined a
very high fraction of radial velocity variable sdB stars, indicating that 
about two thirds of the sdB stars in the field are in close binaries with periods of less 
than 30 days (see also Morales-Rueda et al. \cite{morales}; Napiwotzki et al. \cite{napiwotzki8}). The companions, as far as their nature could be clarified, are mostly M dwarfs or white dwarfs. If the white dwarf companion is sufficiently massive, the merger of the binary 
system might exceed the Chandrasekhar mass and explode as a type Ia supernova. Indeed,  Bill\`{e}res et al. (\cite{billeres00}) and Maxted et al. (\cite{maxted}) discovered KPD~1930$+$2752, a system that might qualify as a SN Ia supernova progenitor (see also Geier et al. \cite{geier}).
 
These discoveries triggered new theoretical evolutionary calculations in the
context of binary population-synthesis to identify the importance of 
various channels of close-binary evolution (Han et al. \cite{han1,han2}), i.e. two phases of 
common-envelope ejection, stable Roche-lobe overflow and white dwarf merger.   
 
\subsection{Outline of the paper}

The purpose of this paper is to clarify the nature of the unseen companions 
for 40 short-period sdB binaries, which comprises about half of the sdB stars in single-lined close 
binary systems with known periods and radial velocity amplitudes.
We assumed tidally locked rotation and made use of the sdBs' gravities and projected rotational velocities. 
 
The paper is structured in two parts. After a short review on close binary 
sdB stars (Sect.~\ref{sec:binaries}), part I (Sects.~\ref{sec:obs} to
\ref{sec:form}) describes the analysis of the sample. 
Besides constraining the mass of the companions and unravelling the
nature of most companions as M dwarfs or typical white dwarfs, it reports the 
discovery of a population of eight unseen compact companions with masses exceeding $0.9\,M_{\rm \odot}$ (in addition to
KPD~1930$+$2752), some of which even exceed 
the Chandrasekhar limit. Accordingly, the latter should be neutron 
stars (NS) or black holes (BH). Even if they were massive white dwarfs, it would be surprising to find
such a large fraction, as massive white dwarfs are rare.  
As no binary system containing an sdB plus a NS/BH is known today, 
we investigate
potential formation scenarios in Sect.~\ref{sec:form} and find it indeed
possible to create such systems through two phases of common envelope evolution. 
 
Our results rest on the assumption of tidally locked rotation. Therefore, part
II of the paper (Sects.~\ref{sec:tidal} and \ref{sec:challenge}) deals with the synchronisation time
scales of sdB stars in close binaries both from a theoretical point of view 
and from the perspective of empirical constraints. 
The general result is that systems with periods
shorter than $1.2\,{\rm d}$ should be synchronised. Empirical evidence is available that
systems with periods below $0.6\,{\rm d}$ are synchronised as is indeed the case for 
the systems with massive companions. 

Although selection effects would favour detection of highly inclined systems,
no such system was found among those binaries with massive companions 
in our sample. This calls for a careful
inspection of alternative explanations (Sect.~\ref{sec:challenge}). There are two aspects to be discussed. 
First, the sdB may not burn helium at all and, thus, is spun up due to ongoing contraction. 
Alternatively the actual evolutionary age of
individual stars may be smaller than appreciated, i.e. the EHB star may just have formed only 
recently and the systems would therefore not be 
synchronised. In Sect.~\ref{sec:summary} we summarise and discuss the results. 
 
\section{Hot subdwarf binaries}\label{sec:binaries}

\begin{figure}[t!]
\begin{center}
	\resizebox{\hsize}{!}{\includegraphics{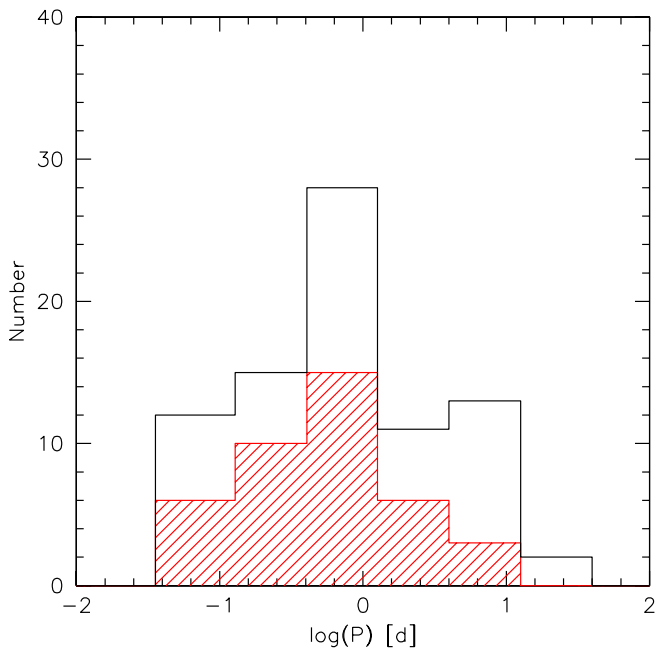}}
	\caption{Period distributions of the 40 binaries in our sample with known orbital parameters (dashed histogram) and all known 81 sdB binaries in the Ritter \& Kolb (\cite{ritter}) catalogue (blank histogram). }
	\label{fig:ritter}
\end{center}
\end{figure}

Several studies aimed at determining the fraction of hot subdwarfs 
residing in close binary systems. Samples of hot subdwarfs have been checked for 
RV variations. The resulting fractions range from 
$39\,\%$ to $78\,\%$ (Green et al. \cite{green4}; Maxted et al. \cite{maxted2};
 Napiwotzki et al. \cite{napiwotzki8}). Several studies were undertaken to 
 determine the orbital parameters of subdwarf binaries 
 (Edelmann et al. \cite{edelmann}; Green et al. \cite{green3}; 
 Morales-Rueda et al. \cite{morales,morales2}; Karl et al. \cite{karl3}). 
 The orbital periods range from $0.07-30\,{\rm d}$ with a peak at 
 $0.5-1.0\,{\rm d}$ (see Fig. \ref{fig:ritter}).  

\subsection{Binary evolution}

For close binary sdBs common envelope ejection is the most probable 
formation channel. In this scenario two main sequence stars of 
different masses evolve in a binary system. The heavier one will first 
reach the red giant phase and fill its Roche lobe. If the mass transfer to the 
companion is dynamically unstable, a common envelope (CE) is formed. 
Due to friction the two stellar cores lose orbital energy, which is deposited 
within the envelope and leads to a shortening of the binary period. 
Eventually the common envelope is ejected and a close binary system is formed, 
which contains a core helium-burning sdB and a main sequence companion. If this star reaches the red giant branch, another common envelope phase is possible 
and can lead to a close binary with a white dwarf companion and an sdB.

If the mass transfer to the companion is dynamically stable, no common envelope
is formed and the primary slowly accretes matter from the secondary. The
companion eventually loses most of its envelope and evolves to an sdB. This
leads to sdB binaries with much larger separation and therefore much longer
orbital periods. Although lots of sdBs have spectroscopically visible main
sequence companions, no radial velocity variable system was detected up to now. Therefore the so called stable Roche lobe overflow (RLOF) channel remains without proof.

Binary evolution also provides a possibility to form single sdB stars via the
merger of two helium white dwarfs (Webbink \cite{webbink}; Iben \& Tutukov 
\cite{iben}). Close He white dwarf binaries are formed as a result of two CE-phases. 
Loss of angular momentum through emission of gravitational radiation will 
cause the system to shrink. Given the initial separation is short enough the 
two white dwarfs eventually merge and if the mass of the merger is high enough, 
core helium burning is ignited and an sdB with very thin hydrogen envelope is 
formed. Recently Politano et al. (\cite{politano}) proposed a new 
evolutionary channel. The merger of a red giant and a low mass main-sequence star during 
a common envelope phase may lead to the formation of a rapidly rotating hot 
subdwarf star. Soker (\cite{soker}) proposed similar scenarios with planetary companions. 
A candidate substellar companion to the sdB star HD\,149382 has been discovered 
recently (Geier et al. \cite{geier5}).

\subsection{SN Ia progenitors}

Double degenerate systems in close orbits 
are viable candidates for progenitors of type Ia supernovae (SN~Ia), which play
 an important role as standard candles for the study of cosmic evolution 
 (e.g. Riess et al. \cite{riess}; Leibundgut \cite{leibundgut}; Perlmutter et 
 al. \cite{perlmutter}). The nature of their progenitors is still under debate 
 (Livio \cite{livio}). The progenitor population provides crucial information 
 for backing the assumption that distant SN~Ia can be used as standard 
 candles like the ones in the local universe.

There is general consensus that only the thermonuclear explosion of a white 
dwarf (WD) is compatible with the observed features of SN~Ia. For this a white 
dwarf has to accrete mass from a close companion to reach the Chandrasekhar 
limit of $1.4 \,M_{\rm \odot}$ (Hamada \& Salpeter \cite{hamada}). According 
to the so-called double degenerate scenario (Iben \& Tutukov \cite{iben}), the 
mass-donating companion is a white dwarf, which eventually merges with the 
primary due to orbital shrinkage caused by gravitational wave radiation. 
A progenitor candidate for the double degenerate scenario must have a total 
mass near or above the Chandrasekhar limit and has to merge in less than a 
Hubble time. Systematic radial velocity (RV) searches for double degenerates 
have been undertaken (e.g. Napiwotzki \cite{napiwotzki2} and references therein)
. The largest of these projects was the ESO SN Ia Progenitor Survey 
(SPY, Napiwotzki et al. \cite{napiwotzki9}). The best known double degenerate 
SN\,Ia progenitor candidate system KPD\,1930$+$2752 has an sdB primary\footnote{The more massive component of a binary is usually defined as the primary. But in most close sdB binaries with unseen companions the masses are unknown and it is not possible decide a priori which component is the most massive one. For this reason we call the visible sdB component of the binaries the primary throughout this paper.}, which 
will become a white dwarf within about $10^{8}\,{\rm yr}$ before the merger occurs in 
about $2\times10^{8}\,{\rm yr}$ (Maxted et al. \cite{maxted}; Geier et al. \cite{geier}).
 Another sdB+WD binary with massive companion has been found recently (Geier 
 et al. \cite{geier4}). 

Most recently Mereghetti et al. (\cite{mereghetti}) showed that in the X-ray 
binary HD\,49798 a very massive ($>1.2\,M_{\rm \odot}$) white dwarf accretes 
matter from the wind of its closely orbiting subdwarf O companion. 
Iben \& Tutukov (\cite{iben94}) predicted that such a system will evolve into a SN Ia when the
primary fills its Roche lobe and transfers mass to the white dwarf to reach the 
Chandrasekhar limit. This makes HD\,49798 a candidate for SN\,Ia progenitor 
for this so called single degenerate scenario.

\subsection{Nature of the companions}

An up-to-date compilation of hot subdwarf binaries with known orbital 
parameters is presented by Ritter \& Kolb (\cite{ritter}) which lists 81 such
systems. 
In general it is difficult to put constraints on the nature of the close
companions of sdB stars. Since most of these binaries are single-lined, only
lower limits to the companion masses could be derived from the stellar mass
functions, which are in general compatible with late main sequence stars of
spectral type M or compact objects like white dwarfs. For single-lined binaries
with longer orbital periods the stellar mass function can help to further
constrain the nature of the unseen companion. Assuming the canonical mass
($0.47\,M_{\rm \odot}$; Han et al. \cite{han1,han2}) for the subdwarf, the
minimum mass of the companion may be  high enough to exclude main sequence 
stars,
 because they would contribute significantly to the flux and therefore appear
  in  the spectra. This mass limit lies near $0.45\,M_{\rm \odot}$ 
 (Lisker et al. \cite{lisker}). 

Twelve sdB binaries have been reported to show eclipses. A combined analysis of the 
light curves and time resolved spectra of these stars allows to derive the 
system parameters as well as the companion types. 
Eight of them have late M companions (see For et al. \cite{for} for a review), while four show shallow variations caused by the eclipse of a white dwarf (Orosz \& Wade \cite{orosz}; Green et al. \cite{green}; Bloemen et al. \cite{bloemen}). 

If close binary stars are double-lined, the mass ratio of the systems can be 
derived from the RV semi-amplitudes of the two components. Until recently, only 
one double-lined He-sdB+He-sdB binary could be analysed 
(Ahmad \& Jeffery \cite{ahmad4}). 

Light variations can help to unravel the nature of the companion by means of the reflection effect and by ellipsoidal
variations, even if there are no eclipses. 
In short period sdB binaries with orbital periods up to about half a day and 
high inclination the hemisphere of a cool main sequence or substellar 
companion directed towards the subdwarf is significantly heated up by the hot 
primary. This leads to a characteristic modulation of the light curve with the
 orbital period, which is a clear indication for an M-star or substellar 
 companion. Such light variations are easily measured in short period binaries 
 with high orbital inclinations. Fourteen sdB+M binaries with 
 this so-called reflection effect are known so far. Since detailed physical 
 models of the reflection effect are not available yet, several free parameters
  have
 to be adjusted to fit the observed light curves. Only very limited constraints
 can therefore be put on the companion masses and radii from an observed 
 reflection effect alone. The absence of a reflection effect can also help to 
 constrain the nature of the unseen companions (Maxted et al. \cite{maxted5}; Shimanskii et al. \cite{shimanskii}). This method works best for binaries with periods of less than $0.5\,{\rm d}$ because otherwise
 the expected reflection effect from an M dwarf companion is hard to detect (Drechsel priv. comm.; Napiwotzki et al. in prep.). The binary JL\,82 shows a very strong reflection, because it is clearly detectable despite the long orbital period of $0.74\,{\rm d}$. What causes the strong variation is not yet understood (Koen \cite{koen2}, see also Sect. \ref{sec:lowmassm}). 

A massive white dwarf companion was identified as companion of an sdB 
(Bill\`{e}res et al. \cite{billeres00}; Maxted et al. \cite{maxted2}; Geier et al. \cite{geier}), which shows a variation in its light curve caused by the tidal distortion of the sdB. Similar signs of ellipsoidal 
 deformation could be detected in five other cases (Orosz \& Wade \cite{orosz};
  O'Toole et al. \cite{otoole}; Geier et al. \cite{geier2}; Koen et al. \cite{koen3}; Bloemen et al. \cite{bloemen}). These stars must have white dwarf companions, because the effect of tidal distortion in the 
  light curve is much weaker than a reflection effect, if present. 

From 81 close binary subdwarfs with known orbital parameters 
(Ritter \& Kolb \cite{ritter}), 13 have bona fide M dwarf companions, while 
7 companions have to be white dwarfs. In another 11 binaries compact 
companions are most likely. One of the binaries has a subdwarf companion. The 
nature of the unseen companions in the remaining 50 binaries could not be 
clarified with the methods described so far. 

Some hot subluminous stars may not be connected to EHB-evolution at all, as
exemplified by HD\,188112 (Heber et al. \cite{heber5}), which was found to be of
too low mass to sustain helium burning in the core. Its atmospheric parameters
place the star below the EHB. An object like HD\,188112 is considered to be a direct 
progenitor of low-mass white dwarfs (Liebert et al., \cite{liebert}), 
which descend from the red giant branch 
and cool down.

\subsection{Rotational properties}

While the rotational properties of blue horizontal branch (BHB) stars both in 
globular clusters and in the field are thoroughly examined 
(see e.g. Behr \cite{behr}), there is no systematic study for EHB stars yet. 
Most of the sdB stars where $v_{\rm rot}\sin{i}$-measurements are available, are slow rotators (Heber et al. \cite{heber2}; Napiwotzki et al. \cite{napiwotzki3}; Edelmann \cite{edelmann}).

The knowledge of the projected rotational velocity, combined with the
gravity determination, allows to derive the mass of single-lined binaries, if the rotation is tidally locked to the orbit. A similar technique has been applied to low-mass X-ray binaries. Kudritzki \& Simon (\cite{kud1}) made use of this method for the first time in the field of hot subdwarfs to constrain the parameters of the sdO binary HD\,49798. Recently, also a few sdB systems have been studied in this way (e.g. Napiwotzki et al. \cite{napiwotzki3}; O'Toole et al. \cite{otoole3}; Geier et al. \cite{geier}, \cite{geier2}, \cite{geier4}). Here we apply this technique to a much larger sample.

\section*{Part I: Quantitative spectral analysis and binary evolution}

Here we present our measurements of projected rotational velocities for 
a sample of 51 radial velocity variable sdBs stars in total. 40 of them are drawn from the Ritter \& Kolb (\cite{ritter}) catalogue (including GD\,687, a system published more recently, Geier et al. \cite{geier4}) and have well determined orbital parameters. 
Eleven additional radial velocity variable sdB stars have also been analysed,
but orbital parameters have not yet been determined.
The main aim is to constrain the masses of the companions under the assumption of tidally locked rotation.

Observations and analysis method are described in
Sects.~\ref{sec:obs} and  \ref{sec:ana}. Surface gravity (Sect.~\ref{sec:atmo}) 
and projected rotational velocities (Sect. \ref{sec:rot}) will be combined with
the mass function to derive companion masses and inclinations. The nature of the companions is discussed  Sect.~\ref{sec:masses}. An evolutionary scenario for the formation of neutron star or black hole companions to sdB stars is proposed in Sect.~\ref{sec:form}.

\section{Observations and Data Reduction \label{sec:obs}}

The first set of UVES spectra were obtained in the course of the 
ESO Supernovae Ia 
Progenitor Survey (SPY, Napiwotzki et al. \cite{napiwotzki9,napiwotzki2}) 
at spectral resolution $R\simeq20\,000-40\,000$ covering 
$3200-6650\,{\rm \AA}$ with two small gaps at $4580\,{\rm \AA}$ and 
$5640\,{\rm \AA}$. Each of the 19 stars were observed at least twice. 
The data reduction is described in Lisker et al. 
(\cite{lisker}). For some of the systems follow-up 
observations with UVES in the same setup were undertaken to derive the orbital 
parameters. These were taken through a narrow slit for better accuracy. 
For the high priority target PG\,1232$-$136 we obtained 60 short 
exposures ($2\,{\rm min}$) with UVES through a very narrow slit ($0.4"$) 
to achieve higher resolution ($R=80\,000$) covering $3770-4980\,{\rm \AA}$ and $5690-7500\,{\rm \AA}$.

High resolution spectra ($R=30\,000$, $4260-6290\,{\rm \AA}$) of 12 known
 close binary subdwarfs have been taken with the HRS fiber spectrograph at the 
 Hobby Eberly Telescope (HET) in the second and third trimester 2007. 
 The spectra were reduced using standard ESO MIDAS routines. 

Another sample of 11 known bright subdwarf binaries was observed with the 
FEROS spectrograph ($R=48\,000$, $3750-9200\,{\rm \AA}$) mounted at the ESO/MPG 
2.2m telescope. The spectra were downloaded from the ESO science archive and 
reduced with the FEROS-DRS pipeline under the ESO MIDAS context in optimum 
extraction mode. 

Three spectra of subdwarf binaries were obtained with the FOCES spectrograph 
($R=30\,000$, $3800-7000\,{\rm \AA}$) mounted at the CAHA 2.2m telescope.
 Three spectra were taken with the HIRES instrument ($R=45\,000$, 
 $3600-5120\,{\rm \AA}$) at the Keck telescope. Two spectra taken with the 
 echelle spectrograph ($R=20\,000$, $3900-8060\,{\rm \AA}$) at the 1.5m 
 Palomar telescope were provided by N. Reid (priv. comm.).

Because a wide slit was used in the SPY survey and the seeing
disk did not always fill the slit, the instrumental profile of some of the UVES spectra was seeing dependent. 
This has to be accounted for to estimate the instrumental resolution. 
The seeing of all single exposures was measured with the DIMM seeing monitor 
at Paranal Observatory and taken from the ESO science archive 
(Sarazin \& Roddier \cite{sarazin}). As a test the seeing was also estimated from 
 the width of the echelle orders perpendicular to the direction of dispersion in some cases 
 and found to be consistent with the DIMM measurements. 
 The errors are considered to be lower than the change of seeing during the 
 exposures (up to $0".2$). 
The resolution of the spectra taken with the fiber spectrographs 
  FEROS, FOCES and HRS was assumed as constant. Changes in the instrumental
   resolution because of temperature variations and for other reasons were 
   considered as negligible.

The single spectra of all programme stars were RV-corrected and co-added in 
order to achieve higher signal-to-noise.

\section{Analysis method \label{sec:ana}}

Since the the programme stars are single-lined spectroscopic binaries, 
no information about the orbital motion of the sdBs' companions is available, 
and thus only their mass functions can be calculated.

 \begin{equation}
 \label{equation-mass-function}
 f_{\rm m} = \frac{M_{\rm comp}^3 \sin^3i}{(M_{\rm comp} +
   M_{\rm sdB})^2} = \frac{P K^3}{2 \pi G}
 \end{equation}

Although the RV semi-amplitude $K$ and the period $P$ are determined by the RV 
curve, the sdB mass $M_{\rm sdB}$, the companion mass $M_{\rm comp}$ and the 
inclination angle $i$ remain free parameters.

In the following analysis we adopt the mass range for sdBs in binaries which 
underwent the common envelope channel given by Han et al. (\cite{han1}, 
\cite{han2}) if no independent mass determinations are available (see 
Sect.~\ref{sec:masses} for details). 

In close binary systems, the rotation of the stars becomes synchronised to their 
orbital motion by tidal forces (see Sect.~\ref{sec:tidal} for a detailed 
discussion). In this case their rotational periods equal the orbital periods 
of the binaries. If the sdB primary is synchronised in this way its rotational 
velocity $v_{\rm rot}$ can be calculated.

\begin{equation}
v_{\rm rot} = \frac{2 \pi R_{\rm sdB}}{P}
\end{equation}

The stellar radius $R$ is given by the mass-radius relation and can be derived,
 if the surface gravity $g$ has been determined.

\begin{equation}
R = \sqrt{\frac{M_{\rm sdB}G}{g}}
\end{equation}

The measurement of the projected rotational velocities $v_{\rm obs}=v_{\rm rot}\,\sin\,i$
 and the surface gravities $g$ therefore allows to constrain the systems' 
 inclination angles $i$. With $M_{\rm sdB}$ as free parameter the mass function
  can be solved and the inclination angle as well as the companion mass can be
   derived. Because of $\sin{i} \leq 1$ a lower limit for the sdB mass is 
   given by

\begin{equation}
\label{eq:minmass}
M_{\rm sdB} \geq \frac{v_{\rm obs}^{2} P^{2}g}{4 \pi^{2}G}
\end{equation}

This method has already been applied to the sdB+WD binaries HE\,1047$-$0436 
(Napiwotzki et al. \cite{napiwotzki3}), Feige\,48
 (O'Toole et el. \cite{otoole3}), KPD\,1930$+$2752 
 (Geier et al. \cite{geier}), PG\,0101$+$039 (Geier et al. \cite{geier2}) and 
 GD\,687 (Geier et al. \cite{geier4}).

There are no signatures of companions visible in the optical spectra of our 
programme stars. Main sequence stars with masses higher than 
$0.45\,M_{\rm \odot}$ could therefore be excluded because 
otherwise spectral features of
 the cool secondary (e.g. Mg\,{\sc i} lines at $\simeq5170\,{\rm \AA}$) 
 would appear in the spectra (Lisker et al. \cite{lisker}) and a flux excess in the 
 infrared would become visible in the spectral energy distribution (Stark \& Wade 
 \cite{stark}; Reed \& Stiening \cite{reed}). 

Another possibility to detect M dwarf or brown dwarf companions are reflection effects in the 
binary light curves. The detection of a reflection effect provides solid 
evidence for the presence of an M dwarf or brown dwarf companion. The non-detection of such a 
modulation can be used to constrain the nature of the companion as well, since a compact object like 
a white dwarf would be too small to contribute significantly to the total flux and cause a detectable reflection effect. But constraining the companion type in this way is problematic for several reasons. First of all, the amplitude of the reflection becomes very small (a few mmag) unless the binary has a short period ($<0.5\,{\rm d}$, Drechsel priv. comm.; Napiwotzki et al. in prep.). Unless the photometry is excellent, such shallow variations over long timescales are not detectable from the ground. Furthermore, the amplitude of the modulation depends on the binary inclination, which is not known in general. An sdB+M binary seen at very low inclination does not show a detectable reflection effect. But most importantly the physics behind the reflection effect itself is poorly understood and one has to use rather crude approximations to derive its amplitude. The most recent detection of a surprisingly strong reflection effect in the long period system JL\,82 (Koen \cite{koen2}) illustrates this.  

Some of our programme stars have already been checked for modulations in their light curves. We consider the lack of a reflection effect as significant constraint, if the orbital period of the binary is shorter than $0.5\,{\rm d}$. In this case the companion should be a compact object. In the case of binaries with longer periods the non-detection of a reflection effect is used as consistency check. 

The atmospheric parameters effective temperature and surface gravity of most 
of our programme stars have been derived from low resolution spectra with 
sufficient accuracy and can be taken from literature in most cases. In order 
to measure projected rotational velocities of sdB stars however, high spectral 
resolution is necessary, because the $v_{\rm rot}\sin{i}$ are small in most 
cases.

\section{Determination of the surface gravity and systematic errors 
\label{sec:atmo}}

Since the precise determination of the atmospheric parameters, especially the
 surface gravity, is of utmost importance for our analysis, this section is 
 devoted to the systematic uncertainties dominating the determination of these 
 parameters. Spectra of sdB stars in the literature were analysed either with metal 
line-blanketed LTE model atmospheres or with NLTE model atmospheres neglecting
 metal line blanketing altogether. As pointed out by Heber et al. 
 (\cite{heber2}), Heber \& Edelmann (\cite{heber4}) and Geier et al. 
 (\cite{geier}), systematic differences between these two approaches are 
 present. Most importantly the gravity scale differs by about $0.05\,{\rm dex}$. 

Most of the atmospheric parameters of our programme stars are taken from 
literature and were derived by fitting LTE or NLTE models (Table \ref{orbit}). The 
adopted errors in $\log{g}$ range from $0.05$ to $0.15$. It is important to 
note that all stars except PG\,1336$-$018, HW\,Vir, PG\,1432$+$159 and PG\,2345$+$318 have been analysed with the same 
grids of LTE and NLTE atmospheres and the same fitting procedure. The error 
in surface gravity starts to dominate the error budget of the derived 
parameters as soon as the error in $v_{\rm rot}\sin{i}$ drops below about 
$1.0\,{\rm kms^{-1}}$ (see Sect.~\ref{sec:rot}). 

\begin{figure}[t!]
\begin{center}
	\resizebox{\hsize}{!}{\includegraphics{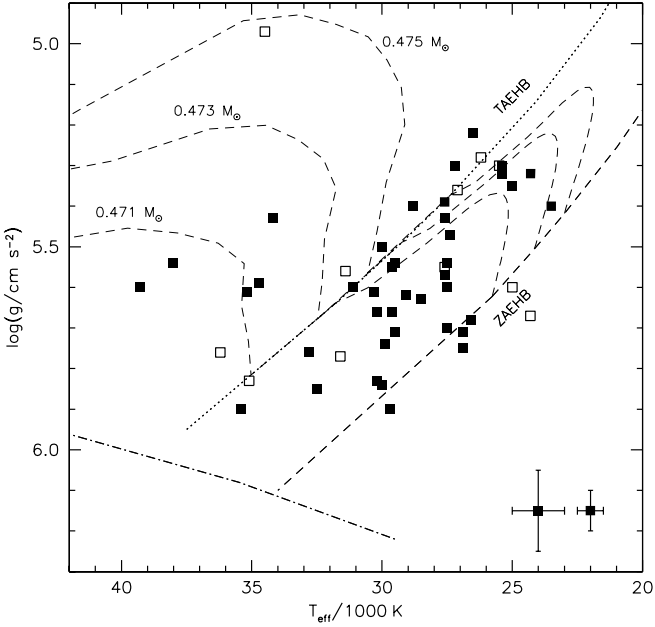}}
	\caption{$T_{\rm eff}-\log{g}$-diagram for the entire sample under study. 
	 The helium main sequence (HeMS) and the EHB band (limited by the zero-age 
         EHB, ZAEHB, and the terminal-age EHB, TAEHB) are superimposed with EHB evolutionary tracks for solar metallicity taken from 
	Dorman et al. (\cite{dorman}) labelled with their masses. Average error bars ($\Delta T_{\rm eff}=500-1000\,{\rm K}$, $\Delta \log{g}=0.05-0.10$) are given in the lower right corner. The filled symbols mark binaries with known orbital parameters (see Table~\ref{orbit}), the open symbols radial velocity variable systems for which orbital parameters are unavailable or uncertain (see Table~\ref{tab:vrotnosol}).}
	\label{fig:tefflogg}
\end{center}
\end{figure}

In cases where no reliable atmospheric parameters could be found in literature, we determined them by fitting LTE models. 
Since the accuracy of the parameters is very much dependent on the higher Balmer 
  lines, a high S/N in this region is necessary. The quality of high resolution 
  spectra obtained with FEROS or FOCES declines toward the blue end. This can cause systematic shifts in the 
  parameter determination (up to $\Delta T_{\rm eff}\simeq2000\,{\rm K}$ and 
  $\Delta \log{g}=0.2$). That is why we chose UVES, HIRES or low resolution 
  spectra to determine the atmospheric parameters if possible. In order to 
  improve the atmospheric parameter determination of TON\,S\,183, 
  BPS\,CS\,22169$-$0001 and $[$CW83$]$\,1735$+$22 we obtained additional 
  medium resolution spectra with WHT/ISIS in August 2009. A medium resolution spectrum of KPD\,1946$+$4340 
  taken with ISIS (Morales-Rueda et al. \cite{morales}) and a low resolution spectrum taken with the B\&C spectrograph 
  mounted at the $2.3\,{\rm m}$ Bok telescope on Kitt Peak (Green priv. comm.) have been fitted with metal-enriched 
  models. 

For the hot stars BPS\,CS\,22169$-$0001, $[$CW83$]$\,1735$+$22 and KPD 1946$+$4340 the NLTE models usually applied gave a strong mismatch for the He\,{\sc ii} line at $4686\,{\rm \AA}$. Using metal line blanketed LTE models of solar composition did not improve the fit. A similar problem was found by O'Toole and Heber (\cite{otoole2}) in their analysis of our programme star CD\,$-$24\,731 (and two other hot sdBs), which is of similarly high temperature. The problem was remedied by using metal enhanced models. Later, the same indication was found for KPD\,1930$+$2752 (Geier et al. \cite{geier}) and AA\,Dor (M\"uller et al. \cite{mueller}). For this reason we used model atmospheres of ten times solar metallicity. Although the atmospheric parameters did not change much, the He\,{\sc ii} line at $4686\,{\rm \AA}$ was matched well in concert with the He\,{\sc i} and hydrogen Balmer lines. 

Only in the case of JL\,82 we had to rely on FEROS spectra. Since the parameters derived from 
  these spectra ($T_{\rm eff}=25\,000\,{\rm K}$, $\log{g}=5.20$) turned out to 
  be very similar to the ones derived from the FEROS spectra of TON\,S\,183 
  ($T_{\rm eff}=26\,000\,{\rm K}$, $\log{g}=5.00$), the systematic shifts 
  ($\Delta T_{\rm eff}=+1500\,{\rm K}$, $\Delta \log{g}=+0.2$) should be 
  similar as well. The parameters of JL\,82 have therefore been corrected for 
  these shifts.

  Results are summarised in Table~\ref{tab:atm} and plotted in
  Fig.~\ref{fig:tefflogg}, where they are compared to canonical models for the EHB band.
  The programme stars populate the EHB band between the zero-age
  (ZAEHB) and the terminal-age EHB (TAEHB). Most of the hottest stars 
  ($>33\,000\,{\rm K}$) are located above the TAEHB and probably have evolved off the
  EHB already. 

\begin{table*}[t!]
\caption{Atmospheric and orbital parameters}\label{tab:atm}
\label{orbit}
\begin{center}
\begin{tabular}{lllllll}
\hline
\noalign{\smallskip}
System & $T_{\rm eff}$ & $\log{g}$ & $P$ & $K$ & $\gamma$ & References\\
       & [K] &  & [d] & [${\rm km\,s^{-1}}$] & [${\rm km\,s^{-1}}$] & \\ 
\noalign{\smallskip}
\hline
\noalign{\smallskip}
PG\,1017$-$086         & 30300 $\pm$ 500 & 5.61 $\pm$ 0.10 & 0.0729938 $\pm$ 0.0000003 & 51.0 $\pm$ 1.7 & -9.1 $\pm$ 1.3 & 14\\
KPD\,1930$+$2752       & 35200 $\pm$ 500 & 5.61 $\pm$ 0.06 & 0.0950933 $\pm$ 0.0000015 & 341.0 $\pm$ 1.0 & 5.0 $\pm$ 1.0 & 7\\
HS\,0705$+$6700       & 28800 $\pm$ 900	& 5.40 $\pm$ 0.10 & 0.09564665 $\pm$ 0.00000039 & 85.8 $\pm$ 3.7 & -36.4 $\pm$ 2.9 & 2\\
PG\,1336$-$018          & 32800 $\pm$ 500 & 5.76 $\pm$ 0.05 & 0.101015999 $\pm$ 0.00000001 & 78.7 $\pm$ 0.6  & -25 & 1,23\\
HW\,Vir         & 28500 $\pm$ 500 & 5.63 $\pm$ 0.05 & 0.115 $\pm$ 0.0008 & 84.6 $\pm$ 1.1 & -13.0 $\pm$ 0.8 & 24,3\\
PG\,1043+760           & 27600 $\pm$ 800 & 5.39 $\pm$ 0.10 &  0.1201506 $\pm$ 0.00000003 & 63.6 $\pm$ 1.4 & 24.8 $\pm$ 1.4 & 13,15\\
BPS\,CS\,22169$-$0001\dag    & 39300 $\pm$ 500 & 5.60 $\pm$ 0.05 & 0.1780 $\pm$ 0.00003 & 14.9 $\pm$ 0.4 & 2.8 $\pm$ 0.3 & 25,5\\
PG\,1432$+$159 & 26900 $\pm$ 1000 & 5.75 $\pm$ 0.15 & 0.22489 $\pm$ 0.00032 & 120.0 $\pm$ 1.4 & -16.0 $\pm$ 1.1 & 21,16\\
PG\,2345$+$318 & 27500 $\pm$ 1000 & 5.70 $\pm$ 0.15 & 0.2409458 $\pm$ 0.000008  & 141.2 $\pm$ 1.1 & -10.6 $\pm$ 1.4 & 22,16\\
PG\,1329$+$159 & 29100 $\pm$ 900 & 5.62 $\pm$ 0.10 & 0.249699 $\pm$ 0.0000002 & 40.2 $\pm$ 1.1 & -22.0 $\pm$ 1.2 & 13,15\\
HE\,0532$-$4503 & 25400 $\pm$ 500 & 5.32 $\pm$ 0.05 & 0.2656 $\pm$ 0.0001 & 101.5 $\pm$ 0.2 & 8.5 $\pm$ 0.1 & 10,19\\
CPD\,$-$64\,481 & 27500 $\pm$ 500 & 5.60 $\pm$ 0.05 & 0.2772 $\pm$ 0.0005 & 23.8 $\pm$ 0.4 & 94.1 $\pm$ 0.3 & 19,5\\
PG\,1101$+$249 & 29700 $\pm$ 500 & 5.90 $\pm$ 0.07 & 0.35386 $\pm$ 0.00006 & 134.6 $\pm$ 1.3 & -0.8 $\pm$ 0.9 & 4,16\\
PG\,1232$-$136 & 26900 $\pm$ 500 & 5.71 $\pm$ 0.05 & 0.3630 $\pm$ 0.0003 & 129.6 $\pm$ 0.04 & 4.1 $\pm$ 0.3 & 25,5\\
Feige\,48 & 29500 $\pm$ 500 & 5.54 $\pm$ 0.05 & 0.376 $\pm$ 0.003 & 28.0 $\pm$ 0.2 & -47.9 $\pm$ 0.1 & 19,20\\
GD\,687    & 24300 $\pm$ 500 & 5.32 $\pm$ 0.07 & 0.37765 $\pm$ 0.00002 & 118.3 $\pm$ 3.4 & 32.3 $\pm$ 3.0 & 11,9\\
KPD\,1946$+$4340 & 34200 $\pm$ 500 & 5.43 $\pm$ 0.10 & 0.403739 $\pm$ 0.0000008 & 167.0 $\pm$ 2.4 & -5.5 $\pm$ 1.0 & 25,15\\
HE\,0929$-$0424 & 29500 $\pm$ 500 & 5.71 $\pm$ 0.05 & 0.4400 $\pm$ 0.0002 & 114.3 $\pm$ 1.4 & 41.4 $\pm$ 1.0 & 10,18\\
HE\,0230$-$4323 & 31100 $\pm$ 500 & 5.60 $\pm$ 0.07 & 0.45152 $\pm$ 0.00002 & 62.4 $\pm$ 1.6 & 16.6 $\pm$ 1.0 & 11,5\\
PG\,1743$+$477 & 27600 $\pm$ 800 & 5.57 $\pm$ 0.10 & 0.515561 $\pm$ 0.0000001 & 121.4 $\pm$ 1.0 & -65.8 $\pm$ 0.8 & 15\\
PG\,0001$+$275 & 25400 $\pm$ 500 & 5.30 $\pm$ 0.10 & 0.529842 $\pm$ 0.0000005 & 92.8 $\pm$ 0.7 & -44.7 $\pm$ 0.5 & 25,5\\
PG\,0101$+$039 & 27500 $\pm$ 500 & 5.53 $\pm$ 0.07 & 0.569899 $\pm$ 0.000001 & 104.7 $\pm$ 0.4 & 7.3 $\pm$ 0.2 & 8\\
PG\,1248$+$164 & 26600 $\pm$ 800 & 5.68 $\pm$ 0.10 & 0.73232 $\pm$ 0.000002 & 61.8 $\pm$ 1.1 & 16.2 $\pm$ 1.3 & 13,15\\
JL\,82 & 26500 $\pm$ 500 & 5.22 $\pm$ 0.10 & 0.73710 $\pm$ 0.00005 & 34.6 $\pm$ 1.0 & -1.6 $\pm$ 0.8 & 25,5\\
TON\,S\,183 & 27600 $\pm$ 500 & 5.43 $\pm$ 0.05 & 0.8277 $\pm$ 0.0002 & 84.8 $\pm$ 1.0 & 50.5 $\pm$ 0.8 & 25,5\\
PG\,1627$+$017 & 23500 $\pm$ 500 & 5.40 $\pm$ 0.10 & 0.8292056 $\pm$ 0.0000014 & 70.10 $\pm$ 0.13 & -54.16 $\pm$ 0.27 & 25,6\\
PG\,1116+301 & 32500 $\pm$ 1000 & 5.85 $\pm$ 0.10 & 0.85621 $\pm$ 0.000003 & 88.5 $\pm$ 2.1 & -0.2 $\pm$ 1.1 & 13,15\\
HE\,2135$-$3749 & 30000 $\pm$ 500 & 5.84 $\pm$ 0.05 & 0.9240 $\pm$ 0.0003 & 90.5 $\pm$ 0.6 & 45.0 $\pm$ 0.5 & 10,18\\
HE\,1421$-$1206 & 29600 $\pm$ 500 & 5.55 $\pm$ 0.07 & 1.188 $\pm$ 0.001 & 55.5 $\pm$ 2.0 & -86.2 $\pm$ 1.1 & 11,18\\
HE\,1047$-$0436 & 30200 $\pm$ 500 & 5.66 $\pm$ 0.05 & 1.21325 $\pm$ 0.00001 & 94.0 $\pm$ 3.0 & 25 $\pm$ 3.0 & 17\\
PG\,0133$+$114 & 29600 $\pm$ 900 & 5.66 $\pm$ 0.10 & 1.23787 $\pm$ 0.000003 & 82.0 $\pm$ 0.3 & -0.3 $\pm$ 0.2 & 15,5\\
PG\,1512$+$244 & 29900 $\pm$ 900 & 5.74 $\pm$ 0.10 & 1.26978 $\pm$ 0.000002 & 92.7 $\pm$ 1.5 & -2.9 $\pm$ 1.0 & 13,15\\
$[$CW83$]$\,1735$+$22 & 38000 $\pm$ 500 & 5.54 $\pm$ 0.05 & 1.278 $\pm$ 0.001 & 103.0 $\pm$ 1.5 & 20.6 $\pm$ 0.4 & 25,5\\
HE\,2150$-$0238 & 30200 $\pm$ 500 & 5.83 $\pm$ 0.05 & 1.321 $\pm$ 0.005 & 96.3 $\pm$ 1.4 & -32.5 $\pm$ 0.9 & 11,18\\
HD\,171858 & 27200 $\pm$ 800 & 5.30 $\pm$ 0.10 & 1.63280 $\pm$ 0.000005 & 87.8 $\pm$ 0.2 & 62.5 $\pm$ 0.1 & 25,5\\
PG\,1716$+$426 & 27400 $\pm$ 800 & 5.47 $\pm$ 0.10 & 1.77732 $\pm$ 0.000005 & 70.8 $\pm$ 1.0 & -3.9 $\pm$ 0.8 & 13,15\\
PB\,7352 & 25000 $\pm$ 500 & 5.35 $\pm$ 0.10 & 3.62166 $\pm$ 0.000005 & 60.8 $\pm$ 0.3 & -2.1 $\pm$ 0.3 & 25,5\\
CD\,$-$24\,731 &  35400 $\pm$ 500 & 5.90 $\pm$ 0.05 & 5.85 $\pm$ 0.003 & 63 $\pm$ 3 & 20 $\pm$ 5 & 19,5\\
HE\,1448$-$0510 & 34700 $\pm$ 500  & 5.59 $\pm$ 0.05 & 7.159 $\pm$ 0.005 & 53.7 $\pm$ 1.1 & -45.5 $\pm$ 0.8 & 10,18\\
PHL\,861 & 30000  $\pm$ 500 & 5.50 $\pm$ 0.05 & 7.44 $\pm$ 0.015 & 47.9 $\pm$ 0.4 & -26.5 $\pm$ 0.4 & 10,18\\
\noalign{\smallskip}
\hline
\end{tabular}
\tablefoot{In the last column references for the atmospheric parameters effective temperature $T_{\rm eff}$ and surface gravity $\log{g}$ (first number) and the orbital parameters period $P$, radial velocity semi-amplitude $K$ and system velocity $\gamma$ (second number) are given separately. If both parameter sets are taken from one source, only one reference number is given. References: $^{1}$Charpinet et al. (\cite{charpinet5}), $^{2}$Drechsel et al. (\cite{drechsel}), $^{3}$Edelmann (\cite{edelmann3}), $^{4}$ Edelmann et al. 
(\cite{edelmann2}), $^{5}$Edelmann et al. (\cite{edelmann}), $^{6}$For et al. (\cite{for2}), $^{7}$Geier et al. (\cite{geier}), $^{8}$Geier et al. (\cite{geier2}), $^{9}$Geier et al. (submitted), $^{10}$Karl et al. (\cite{karl3}), $^{11}$Lisker et al. (\cite{lisker}), $^{12}$Maxted et al. (\cite{maxted4}), $^{13}$Maxted et al. (\cite{maxted2}), $^{14}$Maxted et al. (\cite{maxted3}), 
$^{15}$Morales-Rueda et al. (\cite{morales}), $^{16}$Moran et al. (\cite{moran}), $^{17}$Napiwotzki et al. (\cite{napiwotzki3}), $^{18}$Napiwotzki et al. (in prep.) preliminary results are given in Karl et al. (\cite{karl3}),  $^{19}$O'Toole \& Heber (\cite{otoole2}), $^{20}$O'Toole et al. (\cite{otoole3}), $^{21}$Saffer et al. (\cite{saffer}), $^{22}$Saffer et al. (\cite{saffer2}), $^{23}$Vu\v ckovi\'c et al. (\cite{vuckovic2}), $^{24}$Wood \& Saffer (\cite{wood2}) and this work$^{25}$. \dag The significance of the orbital solution given by Edelmann et al. (\cite{edelmann}) is rather low, but the possible aliases all lie around $0.2\,{\rm d}$.} 
\end{center}
\end{table*}

\section{Projected rotational velocities \label{sec:rot}}

With the gravity at hand, we can derive masses once the projected rotational velocities 
have been measured. This is not an easy task because the sdB stars are known to
be slow rotators. Hence, the broad Balmer and helium lines are ill-suited. 

Sharp metal lines are most sensitive to rotational broadening, in particular for
low velocities, while they tend to be ironed out for fast rotators. 
In order to reach the best accuracy it is necessary to make use 
of as many weak metal lines as possible.

\subsection{Projected rotational velocities from metal lines 
\label{sec:rotlow}}

In order to derive $v_{\rm rot}\,\sin{i}$, we compared the observed spectra 
with rotationally broadened, synthetic line profiles using a semi-automatic 
analysis pipeline. The profiles were computed for the stellar parameters given 
in Table~\ref{orbit} using the LINFOR program (developed by Holweger, Steffen 
and Steenbock at Kiel university, modified by Lemke \cite{lemke}).

\begin{figure*}[t!]
\begin{center}
	\resizebox{\hsize}{!}{\includegraphics{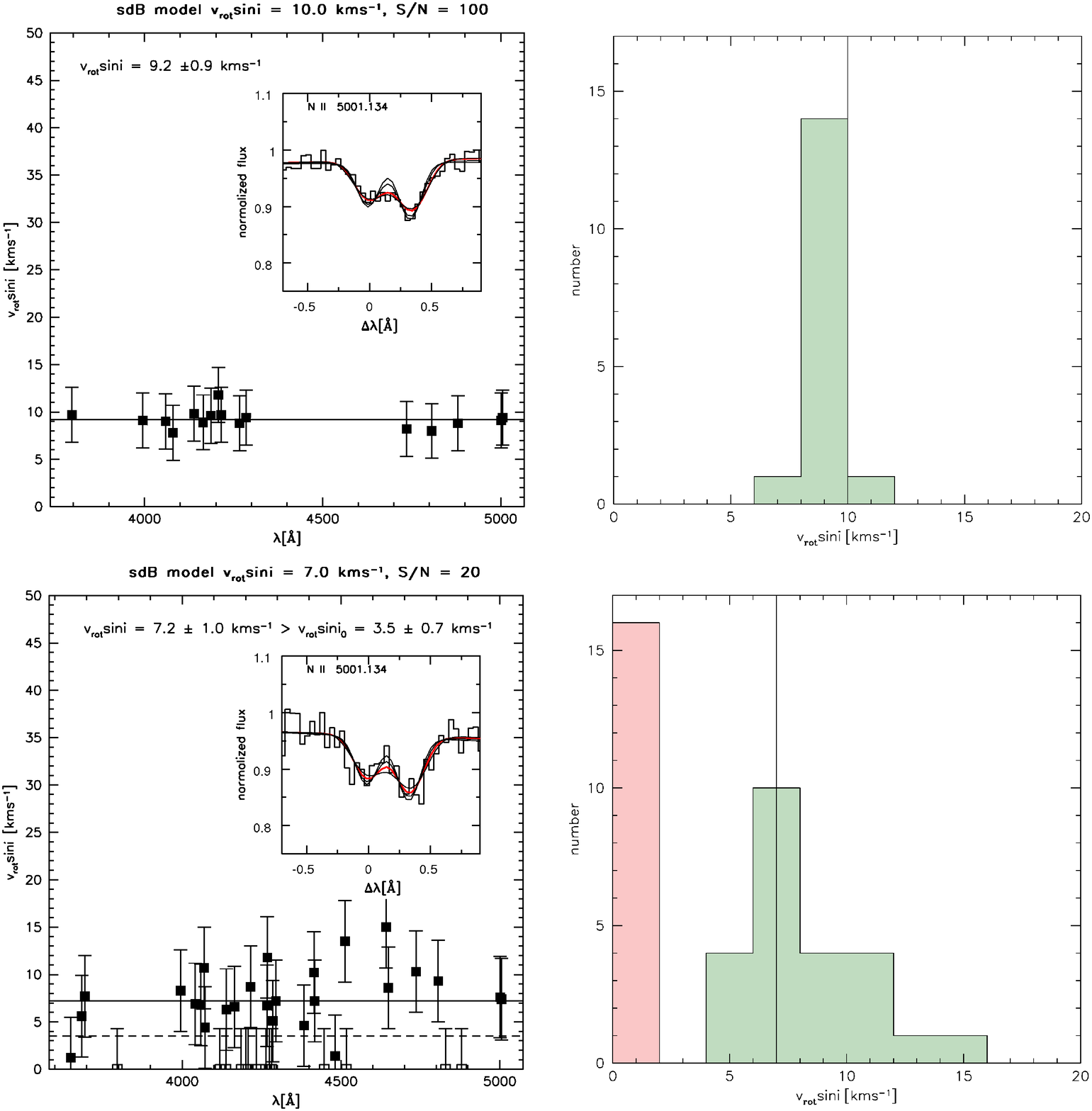}}
	\caption{{\bf Left hand panels:} Numerical simulations. 
	$v_{\rm rot}\sin{i}$ values derived from individual lines are
	 plotted against the wavelength. Standard sdB model spectra with 
	 noise, instrumental and rotational broadening were used for the 
	 calculations. Case A (upper left panel): 
	 $v_{\rm rot}\sin{i} = 10.0\,{\rm km\,s^{-1}}$ and S/N$=100$.
	  The result $9.2\pm0.9\,{\rm km\,s^{-1}}$ is consistent with the true value 
	  within the error margin. The 
	  distribution of individual $v_{\rm rot}\sin{i}$-measurements is shown in the 
	  upper right panel. Case B (lower left panel): 
	  $v_{\rm rot}\sin{i} = 7.0\,{\rm km\,s^{-1}}$ and S/N$=20$. Note that many 
	  lines indicate zero velocity (empty squares). The dashed line corresponds to 
	  the average including the zero values of $3.5\,{\rm km\,s^{-1}}$, which is 
	  systematically lower than the true value. The zero values have to be rejected 
	  in order to obtain the result (solid line): $7.2\pm1.0\,{\rm km\,s^{-1}}$ which 
	  is
	   consistent with the true value within the error margin. {\bf Right hand 
	   panels:} Distribution of individual $v_{\rm rot}\sin{i}$-measurements. The 
	   shaded bin to the left marks the zero values which have to be rejected.}
	\label{normalsdb}
\end{center}
\end{figure*}

\begin{figure}[t!]
\begin{center}
	\resizebox{\hsize}{!}{\includegraphics{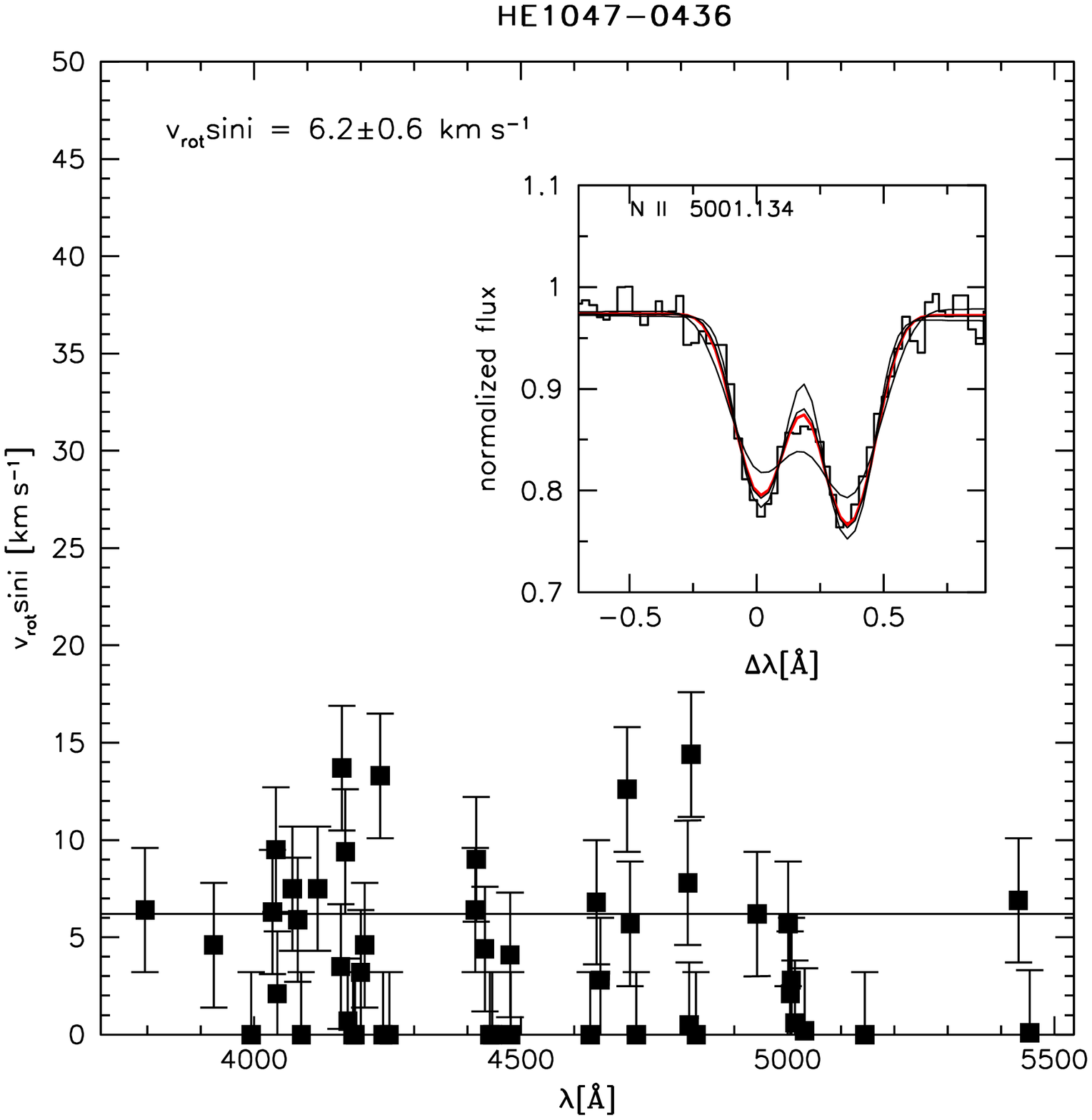}}
	\caption[Rotational broadening fit result for HE\,1047$-$0436.]{Rotational 
	broadening fit result for HE\,1047$-$0436. The measured $v_{\rm rot}\sin{i}$ 
	is plotted against the wavelength of the analysed lines. The solid line 
	corresponds to the average. The inlet shows an example fit of a line doublet. 
	The thick solid line is the best fit $v_{\rm rot}\sin{i}$. The three thin 
	lines correspond to fixed rotational broadenings of $0,5,10\,{\rm kms^{-1}}$.}
	\label{he1047}
\end{center}
\end{figure}

For a standard set of up to 187 unblended metal lines from 24 different ions and with
 wavelengths ranging from $3700$ to $6000\,{\rm \AA}$ a model grid with 
 appropriate atmospheric parameters and different elemental abundances was 
 automatically generated with LINFOR. The actual number of lines used as input 
 for an individual star depended on the wavelength coverage. Due to the 
 insufficient quality of the spectra and the pollution with telluric features 
 in the  regions blueward of $3700\,{\rm \AA}$ and redward of 
 $6000\,{\rm \AA}$ we excluded them from our analysis. A simultaneous fit of 
 elemental abundance, projected rotational velocity and radial velocity was
  then performed separately for every identified line using the FITSB2 
  routine (Napiwotzki et al. \cite{napiwotzki6}). A more detailed description 
  of the line selection and abundance determination will be published in 
  Paper III of this series (Geier et al. in prep.).

\begin{figure}[t!]
\begin{center}
	\resizebox{\hsize}{!}{\includegraphics{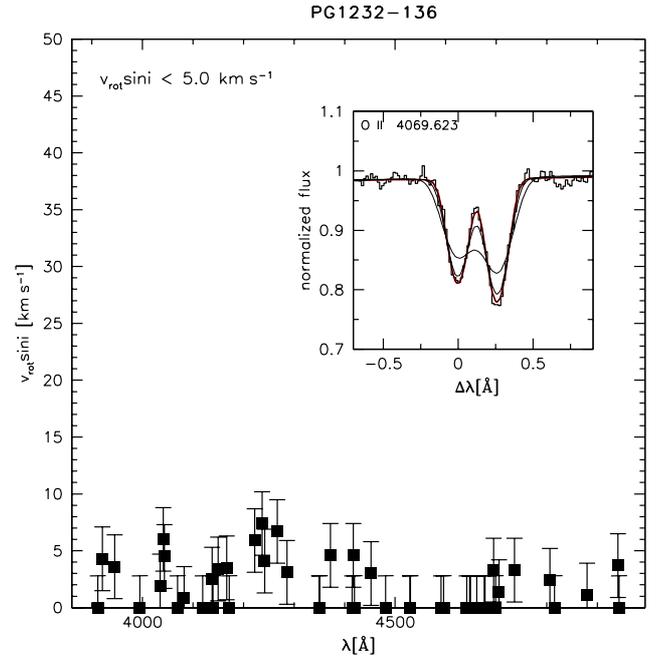}}
	\caption{Rotational broadening fit result for PG\,1232$-$136 
	(see Fig. \ref{he1047}). Despite the high quality of the data no 
	significant $v_{\rm rot}\sin{i}$ could be measured and only an upper 
	limit could be derived.}
	\label{pg1232}
\end{center}
\end{figure}

Ill-suited lines were rejected. This rejection procedure included several 
criteria. First the fitted radial velocity had to be low, because all spectra 
were corrected to zero RV before. Features with high RVs ($>15\,{\rm km\,s^{-1}}$)
 were considered as misidentifications or noise features. Then the fit quality 
 given by the $\chi^2$ had to be comparable to the average. Lines with 
 $\chi^2$-values more than $50\%$ of the average were excluded. A spectral 
 line  was also rejected, if the elemental abundance was lower or higher than 
 the model grid allowed. Equivalent width and depth of the line were measured 
 and compared to the noise to distinguish between lines and noise features. 
 Mean value and statistical error were calculated from all measurements 
 (see Figs. \ref{he1047}$-$\ref{pg1232}). The set of usable lines differs 
 from star to star due to the different atmospheric parameters and chemical 
 compositions. In some cases the line list had to be modified and lines were 
 included or excluded after visual inspection. All outputs of the pipeline 
 have been checked by visual inspection. 

Behr (\cite{behr}) used a similar method to measure the low 
$v_{\rm rot}\sin{i}$ of blue horizontal branch stars from high resolution spectra. 
The errors given in that work are of the same order as the ones given here.

\subsection{Systematic errors in the determination of the projected rotational 
velocity from metal lines \label{sec:rotsystem}}

Since the velocities measured from the metal lines are low, a thorough 
analysis of the errors is crucial. To quantify them, we carried out numerical 
simulations. Synthetic spectra with fixed rotational broadening were computed 
and convolved with the instrumental profile. The standard list of metal lines 
and average sdB parameters ($T_{\rm eff}=30\,000\,{\rm K}$, $\log{g}=5.50$) 
were adopted. Random noise was added to mimic the observed spectra. The 
rotational broadening was measured in the way described above using a grid of
 synthetic spectra for various rotational broadenings and noise levels. 
 As the resolution is seeing dependent for a subset of spectra we also varied 
 the instrumental profile.

Variations in the instrumental profile changed the measured 
$v_{\rm rot}\sin{i}$ by up to $1.0\,{\rm km\,s^{-1}}$ for low S/N and poor 
seeing and about $0.5\,{\rm km\,s^{-1}}$ in case of high S/N and good seeing. 
The noise level caused errors ranging from $2-6\,{\rm km\,s^{-1}}$ per line 
dependent of S/N. Accounting for the number of lines used the error of the average is
 of the order of typically $0.5-3.0\,{\rm km\,s^{-1}}$. A variation of the 
 atmospheric parameters within the derived error limits gives an error of 
 $0.2\,{\rm km\,s^{-1}}$ and is therefore negligible.

We used a standard limb darkening law for the rotational broadening 
independent of wavelength. Berger et al. (\cite{berger}) estimated 
the influence of applying a wavelength dependent limb darkening law on the 
measurements of projected rotational velocities in DAZ white dwarf spectra. 
In the case of the Ca\,{\sc ii} K lines they used, a small difference in the line 
cores was found. Nevertheless, the systematic deviation in $v_{\rm rot}\sin{i}$
 was smaller than $1\,{\rm km\,s^{-1}}$. Because systematic errors caused by this 
 effect would lead to higher real projected rotational velocities than 
 measured, the influence of a wavelength dependent limb darkening law on our 
 results was tested as well. We found the effect to be even lower, because the
  analysed metal lines are much weaker than the Ca\,{\sc ii} K lines used by 
  Berger et al. (\cite{berger}) and the effect becomes more significant for 
  stronger lines. A limb darkening law independent of wavelength is therefore 
  appropriate for our analysis. 

\subsubsection{Individual line fits}

Our numerical experiments included typical numbers of spectral lines ($20-50$)
 as have been used in the analysis spread over the entire wavelength range 
 available ($\simeq3700-6000\,{\rm \AA}$ dependent on the instruments used).
  Fig.~\ref{normalsdb} shows the results of two numerical simulations. The top
   panel displays the result for $v_{\rm rot}\sin{i}=10\,{\rm km\,s^{-1}}$ 
   well above the detection limit and high S/N$=100$. The fitted 
   $v_{\rm rot}\sin{i}$ values for individual lines show small dispersion.

The bottom panel of Fig.~\ref{normalsdb} shows the result for
 $v_{\rm rot}\sin{i}=7\,{\rm km\,s^{-1}}$, which is closer to the detection 
 limit, and low S/N$=20$. Due to the lower S/N individual lines scatter more 
 strongly around the mean. Since negative values of $v_{\rm rot}\sin{i}$ are 
 not possible, the distribution of the measurements is expected to be a 
 truncated Gaussian. As can be seen in the lower right hand panel the 
 distribution doesn't look like a Gaussian, but rather bimodal with many zero 
 measurements. This distribution can be explained, because the truncation of 
 the Gaussian occurs at the detection limit rather than 
 $v_{\rm rot}\sin{i}=0\,{\rm km\,s^{-1}}$. This detection limit is different 
 for each star. It is caused by the thermal broadening of the lines, which 
 scales with $\sqrt{T_{\rm eff}/A}$, $A$ being the atomic weight. The mix of 
 spectral lines used ranges from C ($A=12$) to Fe ($A=56$). The hotter the 
 star, the poorer the result as the number of lines decreases with 
 $T_{\rm eff}$ while the detection limit increases. Other important parameters 
 affecting the detection limit are spectral resolution and the S/N level of
  the spectra.

That is why including the zero values of the bimodal distribution in the 
calculation of the mean would lead to a systematic shift of $v_{\rm rot}\sin{i}$ to 
lower values (see Fig. \ref{normalsdb} lower left panel). For this reason all 
zero values were excluded and the artificial rotational broadening could be
 measured properly. As the lower limit for this method we derived about 
 $v_{\rm rot}\sin{i}>5.0-8.0\,{\rm km\,s^{-1}}$ depending on the resolution of the instrument. 
If more than two thirds of the lines were measured to be zero, this value was adopted as upper limit
for $v_{\rm rot}\sin{i}$. 

As can be seen in the upper panel of Fig.~\ref{normalsdb} the measured mean 
value slightly deviates from the true rotational broadening by 
$0.8\,{\rm km\,s^{-1}}$. Although this deviation is still within the error bars, 
it turned out that such shifts of up to $1\,{\rm km\,s^{-1}}$ can be caused by 
systematic effects. The most likely explanation is that for every individual 
line not only the rotational broadening, but also the elemental abundance is 
fitted. This should affect the $v_{\rm rot}\sin{i}$-distribution and cause a 
deviation from the ideal case of random distribution around the mean. Instead 
of changing the rotational broadening a slightly different elemental abundance 
may lead to a similar $\chi^{2}$-value. Due to this systematic effect a 
minimum $v_{\rm rot}\sin{i}$-error of $1.0\,{\rm km\,s^{-1}}$ is adopted even if 
the statistical error is lower.

Our analysis revealed that the restriction to just a few metal lines in a 
small wavelength range can lead to even higher systematic deviations and that 
it is better to use as many lines as possible scattered over an extended wavelength range 
to measure projected rotational velocities.

There is also an upper limit. With increasing $v_{\rm rot}\sin{i}$ the lines 
are getting broader and broader and eventually cannot be detected any more in 
spectra with S/N typical for our sample. As soon as $v_{\rm rot}\sin{i}$ 
exceeds about $25\,{\rm km\,s^{-1}}$ almost no metal lines can be used unless 
the S/N is much higher than the average of our sample. To measure higher 
projected rotational velocities the Balmer and helium lines must be used as 
described in Sect.~\ref{sec:rothigh}.

\begin{figure}[t!]
\begin{center}
	\resizebox{\hsize}{!}{\includegraphics{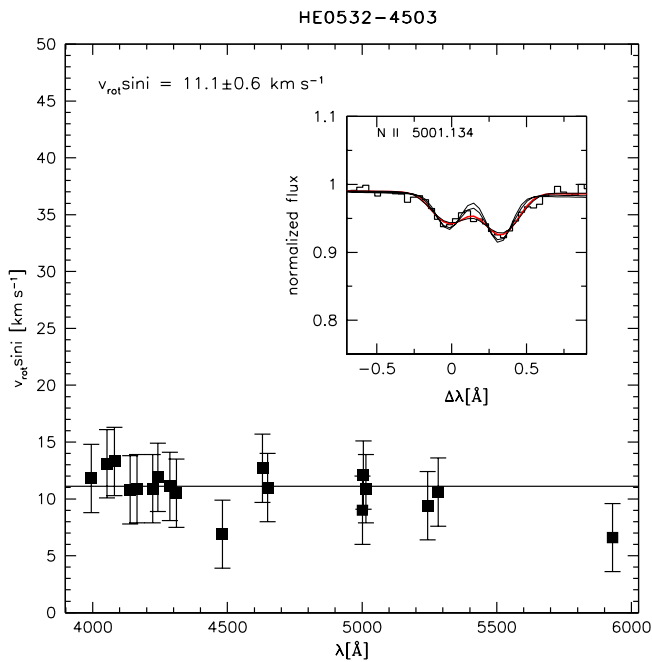}}
	\caption{Rotational broadening fit result for HE\,0532$-$4503 (see 
	Fig. \ref{he1047}).}
	\label{he0532}
\end{center}
\end{figure}

\begin{figure}[]
\begin{center}
	\resizebox{\hsize}{!}{\includegraphics{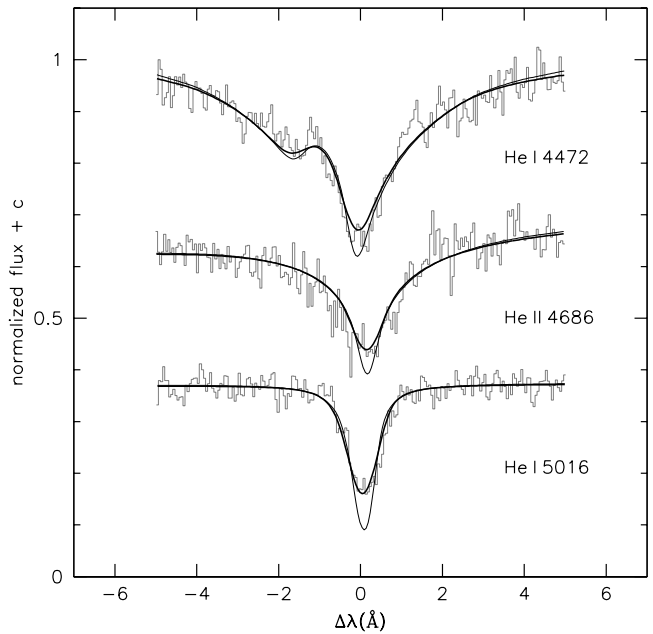}}
	\caption{Selected helium lines of KPD\,1946$+$4340 are plotted against the shift relative to rest wavelengths. The spectrum (histogram) is overplotted with the best fitting rotationally broadened model (strong line). A model without rotational broadening (weak line) is overplotted for comparison.}
	\label{fitkpd1946}
\end{center}
\end{figure}

\subsubsection{Fitting several lines simultaneously}

The FITSB2 routine also allows to fit a lot of lines simultaneously and to use 
different methods of calculating the fitting error (e.g. bootstrapping). 
In principle it is possible to measure the rotational broadening from all 
lines simultaneously and derive the error. But in practice this approach is 
problematic. Fitting up to 25 parameters (24 abundances and 
$v_{\rm rot}\sin{i}$) to more than 50 lines simultaneously and derive the 
error using a bootstrapping algorithm requires a lot of computer power. In 
test calculations we fitted up to nine lines of a synthetic spectrum with 
noise, rotational and instrumental broadening added simultaneously. The
 bootstrap error was consistent with the error we derived with the method 
 described above. Furthermore our error estimate turned out to be slightly 
 higher, which renders our approach more conservative. In the case of very 
 low $v_{\rm rot}\sin{i}$ only some lines remain sensitive to changes in line 
 shape due to rotational broadening. The lower limit that can be reached with 
 the simultaneous approach is therefore higher than what can be detected with 
 the single line approach.

\begin{figure}[t!]
\begin{center}
	\resizebox{\hsize}{!}{\includegraphics{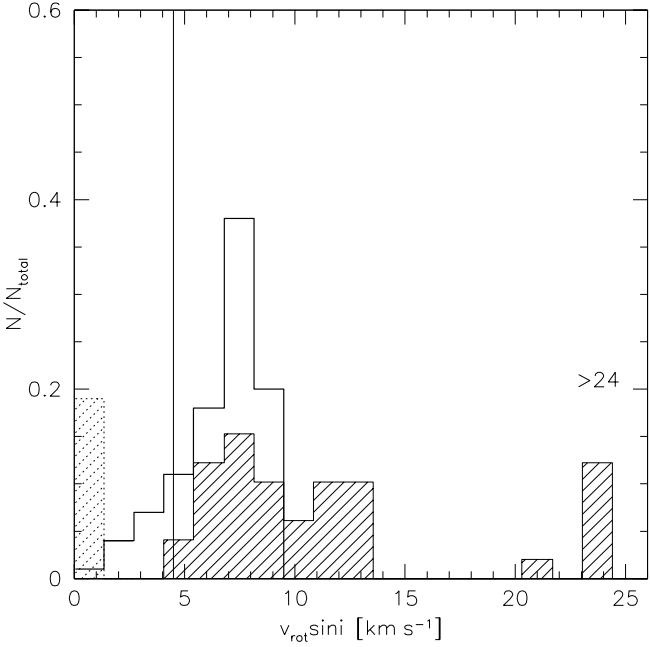}}
	\caption{Shaded histogram showing the distribution of the measured $v_{\rm rot}\sin{i}$ of 51 RV variable sdBs. The blank histogram marks the expected uniform distribution of 
	$v_{\rm rot}\sin{i}$, if the rotational velocity were the same 
	for all stars ($v_{\rm rot}=8.3\,{\rm kms^{-1}}$) and rotation axes 
	were randomly oriented. The solid vertical line at 
	$v_{\rm rot}\sin{i}\simeq5.0\,{\rm kms^{-1}}$ marks the detection 
	limit. All sdBs with lower $v_{\rm rot}\sin{i}$ are stacked into the 
	first bin (dotted histogram). All sdBs with $v_{\rm rot}\sin{i}$ 
	higher than $24\,{\rm kms^{-1}}$ are summed up in the last bin.}
	\label{vrotdistrib_RV}
\end{center}
\end{figure}

\begin{figure}[]
\begin{center}
	\resizebox{\hsize}{!}{\includegraphics{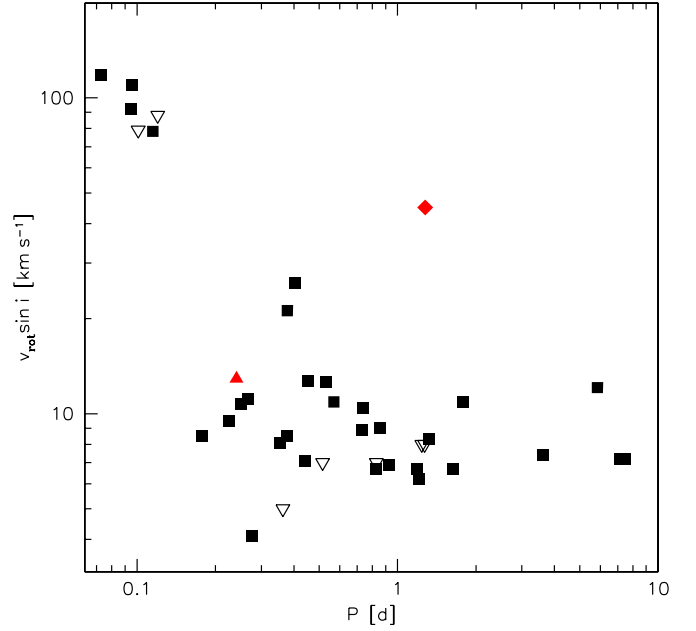}}
	\caption{The measured $v_{\rm rot}\sin{i}$ of 40 RV variable sdBs 
	plotted against the orbital period of the binaries 
	(Tables~\ref{tab:vrotRV}, \ref{orbit}). For seven stars, marked as 
	open inverted triangles, only upper limits were derived. 
	The solid diamond marks $[$CW83$]$\,1735$+$22 that rotates faster than synchronised (see Sect.~\ref{sec:hd188} 
	for a detailed discussion). PG\,2345$+$318 rotates slower than synchronised and is marked with a filled triangle (see Sect.~\ref{sec:age} for a discussion).}
	\label{PvrotI}
\end{center}
\end{figure}

\subsubsection{Orbital smearing}

In the case of binary systems with very short orbital periods 
($0.1-0.2\,{\rm d}$) and high RV amplitudes, the variable Doppler shift of the
 spectral lines during the exposure can lead to a smearing effect, which can 
 be misinterpreted as rotational broadening unless the S/N of the spectra is 
 very high. Orbital smearing is clearly visible in most FEROS spectra of 
 PG\,1232$-$136, which has an orbital period of $0.36\,{\rm d}$ and an 
 RV-semiamplitude of $130\,{\rm km\,s^{-1}}$ (Edelmann et al. \cite{edelmann}). 
 The exposure times of these spectra ranged from $6$ to $30$ minutes. 
 Choosing one single FEROS spectrum with sharp lines obtained at the orbital 
 phase when smearing should be minimal, we derived 
 $v_{\rm rot}\sin{i}=6.2\pm0.8\,{\rm km\,s^{-1}}$ (Geier et al. \cite{geier3}).
  Due to the importance of this object for our conclusions we obtained another 
  60 spectra of PG\,1232$-$136 with UVES at higher resolution ($R=80\,000$). The exposure 
  time of each spectrum was only $2$ minutes. After co-adding all these spectra 
  we constrained $v_{\rm rot}\sin{i}<5.0\,{\rm km\,s^{-1}}$ 
  (see Fig.~\ref{pg1232}). Although the difference between these two results 
  appears to be not very large, it nevertheless illustrates the influence of 
  orbital smearing. 

In the case of the short period ($<0.1\,{\rm d}$) eclipsing sdB+M binary 
HS\,0705$+$6700 with an RV-semiamplitude of $86\,{\rm kms^{-1}}$ the effect 
is much stronger. While Drechsel et al. (\cite{drechsel}) measure 
$v_{\rm rot}\sin{i}=110\pm14\,{\rm kms^{-1}}$ from medium resolution spectra
 with short exposure times ($10-15\,{\rm min}$), we measure 
 $v_{\rm rot}\sin{i}=158\pm12\,{\rm kms^{-1}}$ from a high resolution 
 spectrum taken with HET/HRS and an exposure time of $30\,{\rm min}$. From the
  high resolution data we can only constrain an upper limit of 
  $v_{\rm rot}\sin{i}<170\,{\rm kms^{-1}}$.

Two other stars of our sample (PG\,1336$-$018 and PG\,1043$+$460) may also 
be affected by orbital smearing, if the spectra we used were obtained during 
unfavourable orbital phases. Only upper limits can be given for their 
$v_{\rm rot}\sin{i}$.

\subsubsection{Other systematic errors and their impact on the companion mass determination}

Other possible sources of systematic errors are broadening through 
microturbulence or unresolved pulsations. 
No signi\-ficant microturbulence could be measured which is consistent with the 
analysis of Edelmann et al. (\cite{edelmann4}). Our sample contains six long-period 
pulsating sdBs of V\,1093\,Her type 
(Green et al. \cite{green2}) and four short period pulsators of 
V\,361\,Hya type (Kilkenny et al. \cite{kilkenny}). It has been shown by Telting et al. 
(\cite{telting}) that unresolved high amplitude pulsations with short periods 
can significantly contribute to or even dominate the line broadening. This is 
not a problem for our sample stars, because the pulsation periods of the 
V\,1093\,Her stars are long compared to our exposures times and the amplitudes 
are low. No significant pulsational broadening is expected in the case of the 
short period pulsators Feige\,48 and HE\,0230$-$4323 as well, because the amplitudes of the 
pulsations are low (Reed et al. \cite{reed2}; Charpinet et al. \cite{charpinet2}; Kilkenny et al. \cite{kilkenny2}). 
The line broadening of KPD\,1930$+$2752 and PG\,1336$-$018 is totally dominated by their rotation, 
because the sdBs are spun up by their close companions (Geier et al. 
\cite{geier}; Vu\v ckovi\'c et al. \cite{vuckovic2}).

It has to be pointed out that unresolved pulsations, microturbulence and any 
other unconsidered effect would cause an extra broadening of the lines. 
The true projected rotational velocity would in this case always be lower than 
the one we determined. In this case the derived orbital inclination would also be 
lower and the estimated mass of the unseen companion would be higher (see Sect. \ref{sec:ana}).
Unaccounted systematic effects would therefore lead to higher companion masses. 
This fact is important for the interpretation of the results (see Sect.~\ref{sec:masses}).

\subsection{Projected rotational velocities from hydrogen and 
helium lines \label{sec:rothigh}}

A few sdBs, which reside in close binary systems, are known to be spun up by 
the tidal influence of their companions. The projected rotational velocities 
of these stars are as high as $100\,{\rm km\,s^{-1}}$ (e.g. Drechsel et al. 
\cite{drechsel}; Geier et al. \cite{geier}). 

Rotational broadening irons out the weak metal lines unless the spectra are of
excellent S/N.
However, for higher projected rotational velocities, 
Balmer and helium lines remain the only choice 
to determine $v_{\rm rot}\sin{i}$. Due to 
thermal and pressure broadening Balmer and helium lines are less sensitive to 
rotational broadening than metal lines. From our simulations we derive 
detection limits of $v_{\rm rot}\sin{i}\simeq15\,{\rm km\,s^{-1}}$ for helium 
lines and $v_{\rm rot}\sin{i}\simeq25\,{\rm km\,s^{-1}}$ for the Balmer line 
cores given an S/N$\simeq100$. For lower quality data these limits go up 
significantly. For many of our spectra the Balmer and helium lines are 
insensitive unless $v_{\rm rot}\sin{i}$ exceeds $\simeq 50\,{\rm km\,s^{-1}}$.

To measure the $v_{\rm rot}\sin{i}$ we calculated LTE model spectra with the 
appropriate atmospheric parameters (see Table~\ref{orbit}) and 
performed a simultaneous fit of rotational broadening and helium abundance to 
all usable Balmer line cores and helium lines using the FITSB2 routine 
(Napiwotzki et al. \cite{napiwotzki6}, for an example see Fig.~\ref{fitkpd1946}). All systematic effects discussed in 
the previous section except orbital smearing become negligible in this case. The 
quoted uncertainties are $1\sigma$-$\chi^{2}$-fit errors.

The helium ionisation problem in hot sdBs (see Sect.~\ref{sec:atmo}) caused by neglected metal opacity can affect the measurement of the rotational broadening, if helium lines are used. This became apparent in the analysis of the eclipsing sdOB binary AA\,Dor. While Rauch \& Werner (\cite{rauch}) used metal-free NLTE models and measured $v_{\rm rot}\sin{i}=47\pm5\,{\rm km\,s^{-1}}$ for the He\,{\sc ii} line at $4686\,{\rm \AA}$, Fleig et al. \cite{fleig} measured $v_{\rm rot}\sin{i}=35\pm5\,{\rm km\,s^{-1}}$ by fitting metal line blanketed NLTE models to FUSE spectra. Rucinski (\cite{rucinski}) derived $v_{\rm rot}\sin{i}$ by an analysis of line profile variations during the eclipse and reported a mismatch between the Mg\,{\sc ii} line at $4481\,{\rm \AA}$ and the He\,{\sc ii} line at $4686\,{\rm \AA}$. M\"uller et al. (\cite{mueller}) resolved this conundrum and showed that consistent results ($v_{\rm rot}\sin{i}=30\pm1\,{\rm km\,s^{-1}}$) can be achieved if the appropriate (metal enriched) model atmospheres are used (see Sect.~\ref{sec:atmo}).

To account for this effect we used LTE models with ten times solar metallicity rather than metal-free NLTE models to measure the rotational broadening of the Balmer line cores and helium lines in the two hot sdOBs KPD\,1946$+$4340 (see Fig.~\ref{fitkpd1946}) and $[$CW83$]$\,1735$+$22. While in the case of $[$CW83$]$\,1735$+$22 the $v_{\rm rot}\sin{i}$-values derived with the two different model grids were the same, a significant difference was measured for KPD\,1946$+$4340. The $v_{\rm rot}\sin{i}$ derived with the metal-free models was $42\,{\rm km\,s^{-1}}$ compared to $26\,{\rm km\,s^{-1}}$ with metal-enriched models. 

Due to the fact that KPD\,1946$+$4340 is eclipsing (Bloemen et al. \cite{bloemen}) it is possible to verify that the $v_{\rm rot}\sin{i}$ measured with metal enriched models is fully consistent with the assumption of synchronised rotation (see Sect.~\ref{sec:empirical}).

\subsection{Results \label{sec:close}}

Projected rotational velocities of 46 close binary subdwarfs have been 
measured and supplemented by five measurements taken from literature (Tables~\ref{tab:vrotRV} and ~\ref{tab:vrotnosol}). For 40 systems the orbital parameters are known. In general the projected rotational velocities are small. The other 11 systems are slow rotators, too. These systems can not be analysed further as their mass functions are still unknown. 
  
\begin{table*}[t!]
\caption{Projected rotational velocities for the binary sdB systems from
Table~\ref{tab:atm}.} 
\label{tab:vrotRV}
\begin{center}
\begin{tabular}{lllllllll}
\hline
\noalign{\smallskip}
 System & $T_{\rm eff}$ & $m_{B}$ & S/N & seeing &  $N_{\rm lines}$ &  ${v_{\rm rot}\,\sin\,i}$ & Instrument & Reference\\ 
& [K] & [mag] &  & [arcsec] &  & [${\rm km\,s^{-1}}$] & \\
\noalign{\smallskip}
\hline
\noalign{\smallskip}

PG\,1627$+$017$^{l}$ & 23\,500 & 11.3 & 64 & & 11 & $<$7.0 & HRS &\\
GD\,687 &         24\,300 &      &     &     &     & 21.2 $\pm$ 2.0 & & Geier et al. \cite{geier4}\\
JL\,82$^{l}$ & 25\,000 & 12.2 & 55 & & 57 & 10.4 $\pm$ 1.0 & FEROS & \\
PB\,7352 & 25\,000 & 12.0 & 61 & & 39 & 7.4 $\pm$ 1.0 & FEROS &\\
HE\,0532$-$4503 & 25\,400 & 16.1 &  83 & 0.8 &  18 &  11.1 $\pm$ 1.0 & UVES & \\
PG\,0001$+$275$^{l}$ & 25\,400 & 12.8 & 129 & & 24 & 12.6 $\pm$ 1.0 & FOCES &\\
PG\,1248$+$164 & 26\,600 & 14.4 & 47 & & 13 & 8.9 $\pm$ 1.3 & HRS &\\
PG\,1232$-$136 & 26\,900 & 13.1 & 167 & & 64 & $<$5.0 & UVES &\\
PG\,1432$+$159 & 26\,900 & 13.6 & 50 & & 22 & 9.5 $\pm$ 1.0 & HRS &\\
PG\,1716$+$426$^{l}$ & 27\,400 & 13.7 & 61 & & 24 & 10.9 $\pm$ 1.0 & HRS &\\
PG\,0101$+$039$^{l}$ & 27\,500 &      &    & &   & 10.9 $\pm$ 1.1 & & Geier et al. \cite{geier2}\\
CPD\,$-$64\,481 & 27\,500 & 11.0 & 152 & & 38 & 4.1 $\pm$ 1.0 & FEROS &\\
PG\,2345$+$318 & 27\,500 & 14.4 & 92 & & 21 & 12.9 $\pm$ 1.0 & HRS &\\
PG\,1043$+$760$^{l}$ & 27\,600 & 13.4 & 15 & & H/He & $<$88 & Palomar&\\
PG\,1743$+$477 & 27\,600 & 13.6 & 57 & & 27 & $<$7.0 & HRS &\\
TON\,S\,183 & 27\,600 & 12.4 & 55 & & 57 & 6.7 $\pm$ 1.0 & FEROS &\\
HD\,171858 & 27\,700 & 9.6 & 90 & & 55 & 6.7 $\pm$ 1.0 & FEROS &\\
HW\,Vir & 28\,500 & 10.3 & 130 & & H/He & 78.3 $\pm$ 1.0 & FEROS &\\
HS\,0705$+$6700 & 28\,800 & 14.2 & 28 & & H/He & $<$170 & HRS &\\
                &         &      &    & & H/He &  110 $\pm$ 14 & & Drechsel et al. \cite{drechsel}\\
PG\,1329$+$159 & 29\,100 & 13.3 & 52 & & 26 & 10.7 $\pm$ 1.0 & HRS &\\
Feige\,48$^{s}$ & 29\,500 & 13.1 & 37 & & 36 & 8.5 $\pm$ 1.0 & HIRES &\\
HE\,0929$-$0424 & 29\,500 & 15.4 &  25 & 0.6 &   9 &   7.1 $\pm$ 1.0 & UVES &\\
HE\,1421$-$1206 & 29\,600 & 15.1 &  21 & 0.5 &  18 &   6.7 $\pm$ 1.1 & UVES &\\
PG\,0133$+$114 & 29\,600 & 10.7 & 194 & & 17 & $<$8.0 & FOCES &\\
PG\,1101$+$249 & 29\,700 & 12.5 & 66 & & 24 & 8.1 $\pm$ 1.0 & HIRES &\\
PG\,1512$+$244 & 29\,900 & 13.0 & 87 & & 17 & $<$8.0 & HRS &\\
HE\,2135$-$3749 & 30\,000 & 13.7 &  84 & 1.0 &  53 &  6.9 $\pm$ 1.0 & UVES &\\
PHL\,861 & 30\,000 & 15.1 &  24 & 0.6 &  16 &   7.2 $\pm$ 1.3 & UVES &\\
HE\,1047$-$0436 & 30\,200 & 14.7 &  37 & 0.6 &  37 &   6.2 $\pm$ 1.0 & UVES &\\
HE\,2150$-$0238 & 30\,200 & 15.8 &  27 & 0.8 &  16 &   8.3 $\pm$ 1.5 & UVES &\\
PG\,1017$+$086 & 30\,300  &      &     &     &  H/He & 118 $\pm$ 5  & & Maxted et al. \cite{maxted3}\\
HE\,0230$-$4323$^{s}$ & 31\,100 & 13.8 &  59 & 0.9 &  40 &  12.7 $\pm$ 1.0 & UVES &\\
PG\,1336$-$018$^{s}$ & 31\,300 & 14.0 & 40 & & H/He & $<$79.0 & FEROS &\\
PG\,1116$+$301 & 32\,500 & 14.3 & 42 & & 8 & 9.0 $\pm$ 1.7 & HRS &\\
KPD\,1946$+$4340 & 34\,200 & 14.1 & 55 & & H/He & 26.0 $\pm$ 1.0 & HRS &\\
HE\,1448$-$0510 & 34\,700 & 15.0 &  27 & 0.6 &   8 &   7.2 $\pm$ 1.7 & UVES &\\
KPD\,1930$+$2752$^{s}$ & 35\,200 &      &     &     &     & 92.3 $\pm$ 1.5 & & Geier et al. \cite{geier}\\
CD\,$-$24\,731 & 35\,400 & 11.6 & 42 & &   8 & 12.1 $\pm$ 1.7  & FEROS &\\
$[$CW83$]$\,1735$+$22 & 38\,000 & 11.5 & 230 & & H/He & 44.0 $\pm$ 1.0 & FOCES &\\ 
BPS\,CS\,22169$-$0001 & 39\,300 & 12.6 & 109 & & 5 & 8.5 $\pm$ 1.5 & FEROS &\\
\hline
\\
\end{tabular}
\tablefoot{For binaries with high ${v_{\rm rot}\,\sin\,i}$ 
helium lines and Balmer line cores (H/He) are used instead of metal lines. 
The average seeing is only given if the spectra were obtained with a wide 
slit in the course of the SPY survey. In all other cases the seeing should not 
influence the measurements. $^{c}$Companion visible in the spectrum. 
$^{s}$Pulsating subdwarf of V\,361\,Hya type. $^{l}$Pulsating subdwarf of
 V\,1093\,Her type.}
\end{center}
\end{table*}

\begin{table*}[t!]
\caption{Projected rotational velocities of radial velocity variable sdBs, for
which orbital parameters are unavailable or uncertain.} 
\label{tab:vrotnosol}
\begin{center}
\begin{tabular}{llllllll}
\hline
\noalign{\smallskip}
 System & $T_{\rm eff}$ & $m_{B}$ & S/N & seeing &  $N_{\rm lines}$ &  ${v_{\rm rot}\,\sin\,i}$ & Instrument\\ 
& [K] & [mag] &  & [arcsec] &  & [${\rm km\,s^{-1}}$] & \\
\noalign{\smallskip}
\hline
\noalign{\smallskip}

HE\,2208$+$0126 & 24\,300 & 15.6 &  24 & 0.8 &  15 & $<$5.0 & UVES\\
TON\,S\,135 & 25\,000 & 13.1 &  47 & &  35 &  6.6 $\pm$ 1.0  & FEROS\\
HE\,2322$-$4559$^{c}$ & 25\,500 & 15.5 &  23 & 0.7 &  16 &  10.9 $\pm$ 1.1 & UVES\\
HS\,2043$+$0615 & 26\,200 & 16.0 &  22 & 1.3 &  26 &  12.3 $\pm$ 1.1 & UVES\\
HE\,1309$-$1102$^{c}$ & 27\,100 & 16.1 &  7 & 0.6 &  7 &  7.6 $\pm$ 2.3 & UVES\\
HS\,2357$+$2201 & 27\,600 & 13.3 &  29 & 0.7 &  26 & 6.1 $\pm$ 1.1 & UVES\\
HS\,2359$+$1942 & 31\,400 & 14.4 &  14 & 0.6 &  26 &  $<$ 5.0 & UVES\\
PG\,1032$+$406 & 31\,600 & 10.8 & 20 & & H/He & $<$34 & Palomar\\
HE\,1140$-$0500$^{c}$ & 34\,500 & 14.8 &  18 & 0.9 &   5 &   5.2 $\pm$ 2.7 & UVES\\
HS\,1536$+$0944$^{c}$ & 35\,100 & 15.6 &  19 & 1.1 &  15 &  12.2 $\pm$ 1.6 & UVES\\
HE\,1033$-$2353 & 36\,200 & 16.0 &  13 & 0.6 &   7 &   9.3 $\pm$ 2.3 & UVES\\
\hline
\\
\end{tabular}
\tablefoot{The average seeing is only given if the spectra were obtained with a wide slit in the course of the SPY survey. In all other cases the seeing should not influence the measurements. Atmospheric parameters are taken from Lisker et al. (\cite{lisker}) except TON\,S\,135 (Heber \cite{heber1}) and PG\,1032$+$406 (Maxted et al. \cite{maxted2}). $^{c}$Companion visible in the spectrum.}
\end{center}
\end{table*}

The projected rotational velocities of HE\,1047$-$0436 and Feige\,48 have 
been measured by Napiwotzki et al. (\cite{napiwotzki3}) and O'Toole et al. 
(\cite{otoole3}) using a technique similar to the one described here, but 
restricted to just a few metal lines. Napiwotzki et al. (\cite{napiwotzki3}) 
derived an upper limit of $v_{\rm rot}\sin{i}=4.7\,{\rm km\,s^{-1}}$ for 
HE\,1047$-$0436. Our measurement of 
$6.2\pm0.6\,{\rm km\,s^{-1}}$ is just slightly higher 
(see Fig.~\ref{he1047}). While O'Toole et al. (\cite{otoole3}) give an upper 
limit of $v_{\rm rot}\sin{i}=5\,{\rm km\,s^{-1}}$ for Feige\,48 we derive 
$8.5\pm1.5\,{\rm km\,s^{-1}}$. 

Fig.~\ref{PvrotI} shows the measured $v_{\rm rot}\sin{i}$ plotted against the 
orbital periods of the binaries. A trend is clearly visible: The longer the 
orbital period of the systems, the lower the measured $v_{\rm rot}\sin{i}$. 
While the short period systems ($\simeq0.1\,{\rm d}$) were spun up by their 
close companions and have high $v_{\rm rot}\sin{i}$ up to
 $\simeq100\,{\rm km\,s^{-1}}$, the mean $v_{\rm rot}\sin{i}$ decrease to 
 below $10\,{\rm km\,s^{-1}}$ as the periods increase to $\simeq1.0\,{\rm d}$.
  For orbital periods exceeding $\simeq1.0\,{\rm d}$, the
   $v_{\rm rot}\sin{i}$-values scatter around the average 
   $v_{\rm rot}=8.3\,{\rm km\,s^{-1}}$ for single sdB stars 
   (Geier et al. \cite{geier3}). We conclude that tidal forces do not 
   influence the rotation of sdBs for orbital periods considerably longer than 
   one day.

As can be seen in  Fig.~\ref{vrotdistrib_RV} the 
$v_{\rm rot}\sin{i}$-distribution of the RV variable sdBs (Tables~\ref{tab:vrotRV}, \ref{tab:vrotnosol})
differs from the uniform distribution of the single stars 
 (Geier et al. \cite{geier3}), the rotational properties of the full sample 
 of single sdB stars will be presented in paper II of this series by Geier et 
 al. (in prep.). A large fraction of binary sdBs exceeds the derived maximum 
 $v_{\rm rot}=8.3\,{\rm km\,s^{-1}}$ significantly. The most likely reason for 
 this is tidal interaction with the companions. 

\section{Constraining masses, inclinations and the nature of the unseen
companions}\label{sec:masses}

Having determined the projected rotational velocity we are in a position to 
derive the companion mass as a function of the sdB mass as described in 
Sect.~\ref{sec:ana}. 

\begin{figure}[t!]
\begin{center}
	\resizebox{\hsize}{!}{\includegraphics{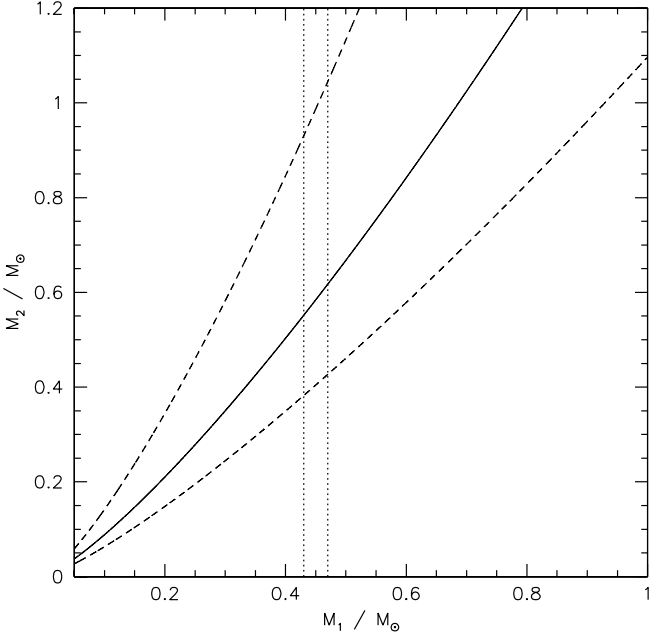}}
	\caption{Mass of the sdB primary CPD\,$-$64\,481 plotted against the mass of the unseen
	 companion. The companion mass error is indicated by the dashed lines. 
	 The mass range of the CE ejection channel (Han et al. \cite{han1}) is 
	 marked with dotted vertical lines.}
	\label{cpdm64481_mass}
\end{center}
\end{figure}

From 40 sdB binaries, for which all necessary parameters have been determined,
 31 could be solved consistently under the assumption of tidally locked rotation.
  Two examples are shown in Figs. 
\ref{cpdm64481_mass} and \ref{he0532_mass}. 
 Derived inclinations, subdwarf masses and the allowed masses for the 
 companions are given in Table~\ref{compmasses}.
 
If the sdB mass could not be constrained with other methods (e.g. from
photometry, see Table~\ref{compmasses}), the theoretically predicted mass range 
was taken from Han et al. (\cite{han1}, \cite{han2}). For the common envelope 
ejection channels, which are the only plausible way of forming sdBs in close 
binary systems, the possible masses for the sdBs range from 
$0.38\,M_{\rm \odot}$ to $0.47\,M_{\rm \odot}$. Since in all simulation sets 
of Han et al. (\cite{han1}, \cite{han2}) the mass distribution 
shows a very prominent peak at $0.43-0.47\,M_{\rm \odot}$ this mass range is 
the most likely one. 

The choice of the adopted sdB mass range is backed up by recent mass
determinations via asteroseismology of 
short-period pulsating sdBs. Fontaine et al. (\cite{fontaine}) showed the 
mass distribution of 12 of these objects, which is in good agreement with 
the predicted distribution by Han et al. (\cite{han1}, \cite{han2}). 
Consistent with theory no star of this small sample has a mass much lower 
than $0.4\,M_{\rm \odot}$. The few sdB masses, that could be constrained by 
analyses of eclipsing binary systems also range from $0.38\,M_{\rm \odot}$ 
to $0.5\,M_{\rm \odot}$ (see e.g.~Sect.~\ref{sec:lowmassm} and For et al. 
\cite{for}). 

Hence we adopt $0.43-0.47\,M_{\rm \odot}$ as the mass range 
for the sdBs in the binary systems we studied, if there is no independent 
mass determination either from binary light curve analysis or asteroseismology. 

If the derived minimum sdB mass assuming a sychronised orbit (see Equation~\ref{eq:minmass}) exceeds 
this reasonable mass range ($M_{\rm sdB} \gg 1\,M_{\rm \odot}$) the sdB primary spins 
faster than synchronised and no consistent solution can be found.
This is the case for 9 binaries from our sample. 
Most of these systems have orbital periods exceeding $1.2\,{\rm d}$, 
where we find that synchronisation is no longer established 
(see Sect.~\ref{sec:tidal}). It has to be pointed out that only 
subdwarfs rotating faster than synchronised can be identified in this way.
 If an sdB should rotate slower than synchronised, one would always get an 
 apparently consistent, but incorrect solution, which overestimates the 
 companion mass (see Sect.~\ref{sec:age}). For {\bf PG\,0133$+$114} there is some doubt whether
 the star is synchronised or not as the minimum mass for the sdB is 
$0.51\,{M_{\rm \odot}}$ at the upper end of the predicted mass range for core 
helium-burning objects and its period is rather long ($1.24\,{\rm d}$). 
The minimum companion mass would be $0.38\,\,{M_{\rm \odot}}$, while the statistically
most likely one ($i=52^{\circ}$) $0.48\,M_{\rm \odot}$, indicating it is a white dwarf, if the
system is synchronised.

The nature of the companion was deduced unambiguously for most of the remaining stars (except five)
from the masses and additional information. The companions to {\bf PG\,1248$+$164}\footnote{A light curve of this star has been taken by Maxted et al. (\cite{maxted5}). No variability could be detected. Although the orbital period is rather long ($0.73\,{\rm d}$) and a reflection effect therefore shallow, the companion may be a low-mass WD rather than an M dwarf.}, 
{\bf HE\,1421$-$1206}, {\bf Feige~48}\footnote{The mass of the pulsating subdwarf Feige\,48 
has been determined in an asteroseismic analysis (van Grootel et al. 
\cite{vangrootel}) to $0.52\,{M_{\rm \odot}}$. The corresponding companion 
mass is $0.27\,{M_{\rm \odot}}$. Therefore the nature of the unseen 
companion remains unclear. 
It may be a low mass white dwarf as well as a late M dwarf.
Due to the derived very low inclination and 
the presence of short period pulsations, a reflection effect or ellipsoidal
variations are probably too small to be detectable.}, and 
{\bf HE\,2135$-$3749} could be either main sequence stars or white dwarfs 
because their
masses are lower than $0.45\,M_{\rm \odot}$. 

We shall describe the results for three groups of companion stars. 
 Starting with sdBs orbited by low mass dwarf companions, we proceed to the 
 systems with white dwarf companions of normal masses. Finally we discuss the 
 group of binaries that contain massive
 compact companions exceeding $0.9\,M_{\rm \odot}$, because such systems
 are of particular interest, e.g. as potential SN Ia progenitors.
 This includes KPD\,1930+2752, the most massive white dwarf companion to an sdB
 star known so far. 
 
 \begin{table*}[t!]
\caption{Derived inclination angles, companion masses and likely
 nature of the companions.} 
\label{compmasses}
\begin{center}
\begin{tabular}{llllllll}
\hline
\noalign{\smallskip}
System & $P^{*}$ &  $M_{\rm sdB}$ & $i$ & $M_{\rm comp}$ & $i_{\rm max}$ & $M_{\rm comp, min}$ & Companion \\
       & [d] & [$M_{\rm \odot}$] & [deg] & [$M_{\rm \odot}$] & [deg] & [$M_{\rm \odot}$] & \\ 
\noalign{\smallskip}
\hline
\noalign{\smallskip}
PG\,1017$-$086$^{12}$ & 0.07 & $>$0.47 & $<$73 & $>$0.06 & & & MS/BD$^{r}$ \\
KPD\,1930$+$2752$^{6}$ & 0.10 & $0.47_{-0.02}^{+0.05}$ & $77_{-4}^{+4}$ & $0.94_{-0.03}^{+0.02}$ & & & WD$^{el}$ \\
HS\,0705$+$6700$^{3}$ & 0.10 & 0.48 & $65_{-16}^{+25}$ & $0.15_{-0.03}^{+0.05}$ & & & MS$^{r,ec}$ \\
PG\,1336$-$018$^{2,16}$  & 0.10 & 0.459 & $<$90 & $>$0.12 & & & MS$^{r}$ \\
HW\,Vir$^{4}$ & 0.12 & 0.53 & $75_{-10}^{+15}$ & $0.155_{-0.015}^{+0.015}$ & & & MS$^{r,ec}$ \\
PG\,1043+760$^{13}$ & 0.12 & & $<$78 & $>$0.10 & 90 & 0.06 & WD$^{n}$ \\
BPS\,CS\,22169$-$0001$^{14}$ & 0.18 & & $9_{-2}^{+2}$ & $0.19_{-0.06}^{+0.07}$ & 13 & 0.09 & MS$^{r}$ \\
PG\,1432$+$159$^{12}$ & 0.22 & & $16_{-3}^{+5}$ & $2.59_{-1.10}^{+2.01}$ & 25 & 0.92 & NS/BH$^{n}$ \\
PG\,2345$+$318$^{2}$ & 0.24 & & & & & & WD$^{ec}$ not synchronised\\
PG\,1329$+$159$^{12}$ & 0.25 & & $17_{-2}^{+4}$ & $0.35_{-0.10}^{+0.10}$ & 26 & 0.16 & MS$^{r}$ \\
HE\,0532$-$4503$^{11}$ & 0.27 & & $14_{-2}^{+2}$ & $3.00_{-0.92}^{+0.94}$ & 19 & 1.27 & NS/BH$^{f}$ \\
CPD\,$-$64\,481 & 0.28 & & $7_{-2}^{+2}$ & $0.62_{-0.24}^{+0.42}$ & 11 & 0.24 & WD \\
PG\,1101$+$249 & 0.35 & & $26_{-4}^{+6}$ & $1.67_{-0.58}^{+0.77}$ & 40 & 0.68 & WD/NS/BH $^{f}$ \\
PG\,1232$-$136 & 0.36 & & $<$14 & $>$6.00 & 17 & 3.58 & BH$^{f}$ \\
Feige\,48$^{15}$ & 0.38 & 0.52 & $17_{-2}^{+3}$ & $0.27_{-0.04}^{+0.06}$ & & & MS/WD \\
GD\,687$^{5,7}$ & 0.38 & & $39_{-6}^{+6}$ & $0.71_{-0.21}^{+0.22}$ & 63 & 0.32 & WD$^{f}$ \\
KPD\,1946$+$4340$^{1}$ & 0.40 & & $71_{-15}^{+19}$ & $0.67_{-0.08}^{+0.18}$ & 90 & 0.58 & WD$^{el,ec}$\\
HE\,0929$-$0424$^{11}$ & 0.44 & & $23_{-4}^{+5}$ & $1.82_{-0.64}^{+0.88}$ & 34 & 0.73 & WD/NS/BH$^{f}$ \\
HE\,0230$-$4323$^{9}$ & 0.45 & & $39_{-5}^{+8}$ & $0.30_{-0.07}^{+0.07}$ & 61 & 0.15 & MS$^{r}$ \\
PG\,1743$+$477 & 0.52 & & $<$27 & $>$1.66 & 32 & 1.00 & NS/BH$^{f}$ \\
PG\,0001$+$275 & 0.53 & & $31_{-4}^{+7}$ & $0.79_{-0.23}^{+0.26}$ & 48 & 0.37 & WD \\
PG\,0101$+$039$^{8}$ & 0.57 & & $40_{-6}^{+9}$ & $0.72_{-0.20}^{+0.20}$ & 64 & 0.33 & WD$^{el,n}$ \\
PG\,1248$+$164 & 0.73 & & $52_{-12}^{+25}$ & $0.27_{-0.08}^{+0.10}$ & 90 & 0.12 & MS/WD \\
JL\,82$^{10}$ & 0.74 & & $33_{-5}^{+8}$ & $0.21_{-0.06}^{+0.06}$ & 51 & 0.10 & MS$^{r}$ \\
TON\,S\,183 & 0.83 & & $30_{-5}^{+7}$ & $0.94_{-0.31}^{+0.39}$ & 47 & 0.40 & WD$^{f}$ \\
PG\,1627$+$017 & 0.83 & & $<$34 & $>$0.50 & 45 & 0.32 & WD \\
PG\,1116$+$301 & 0.86 & & 90 & $0.48_{-0.21}^{+0.00}$ & 90 & 0.27 & WD \\
HE\,2135$-$3749 & 0.92 & & $67_{-16}^{+13}$ & $0.41_{-0.12}^{+0.13}$ & 90 & 0.29 & MS/WD \\
HE\,1421$-$1206 & 1.19 & & $57_{-14}^{+33}$ & $0.27_{-0.08}^{+0.10}$ & 90 & 0.16 & MS/WD \\
HE\,1047$-$0436 & 1.21 & & $62_{-10}^{+28}$ & $0.53_{-0.14}^{+0.15}$ & 90 & 0.28 & WD \\
PG\,0133$+$114 & 1.24 & $>0.51$ & 90 & $>0.38$ & & & MS/WD/not synchronised? \\
PG\,1512$+$244 & 1.27 & & & & & & not synchronised? \\
$[$CW83$]$\,1735$+$22 & 1.28 & & & & & & not synchronised \\
HE\,2150$-$0238 & 1.32 & & &  & & & not synchronised \\
HD\,171858 & 1.63 & & $58_{-14}^{+32}$ & $0.60_{-0.19}^{+0.25}$ & 90 & 0.37 & WD \\
PG\,1716$+$426 & 1.78  & & & & & & not synchronised \\
PB\,7352 & 3.62 & & & & & & not synchronised \\
CD\,$-$24\,731 & 5.85 & & & & & & not synchronised \\
HE\,1448$-$0510 & 7.16 & & & & & & not synchronised \\
PHL\,861 & 7.44 & & & & & & not synchronised \\
\noalign{\smallskip}
\hline\\
\end{tabular}
\tablefoot{If the sdB mass couldn't be constrained with 
 other methods the theoretically predicted mass range of 
 $0.43-0.47\,{M_{\rm \odot}}$ was taken from Han et al. (\cite{han1,han2}).
  The minimum masses of the companions and maximum inclinations of the 
  binaries were calculated for the lowest possible sdB mass 
  ($0.3\,{M_{\rm \odot}}$, Han et al. \cite{han1,han2}). $^{*}$The 
  orbital periods given here are rounded to the second decimal place. The 
  accurate values are given in Table \ref{orbit}. Additional constraints to clarify the nature of the unseen companions: $^{r}$The detection of a reflection effect from a cool MS/BD or a $^{n}$non-detection to exclude 
   this option. The presence of eclipses$^{ec}$ or ellipsoidal 
   deformations$^{el}$ in the light curves. No signatures of a main-sequence companion 
   within the given mass range are visible in the flux distribution or in the
    spectrum$^{f}$. These informations are taken from $^{1}$Bloemen et al. (\cite{bloemen}), $^{2}$Charpinet et al. 
    (\cite{charpinet5}),  $^{3}$Drechsel et al. (\cite{drechsel}), $^{4}$Edelmann 
    (\cite{edelmann3}), $^{5}$Farihi et al. (\cite{farihi}), $^{6}$Geier et al. 
    (\cite{geier}), $^{7}$Geier et al. (\cite{geier4}), $^{8}$Geier et al. 
    (\cite{geier2}), $^{9}$Koen (\cite{koen}), $^{10}$Koen (\cite{koen2}), 
    $^{11}$Lisker et al. (\cite{lisker}), $^{12}$Maxted et al. (\cite{maxted3}),
     $^{13}$Maxted et al. (\cite{maxted5}), $^{14}$\O stensen (priv. comm.),
      $^{15}$van Grootel et al. (\cite{vangrootel}) and $^{16}$Vu\v ckovi\'c et 
      al. (\cite{vuckovic2}).}
\end{center}
\end{table*} 
  
\subsection{Late main sequence stars and a potential brown dwarf \label{sec:lowmassm}}

{\bf PG\,1017$-$086} is the sdB binary with the shortest orbital period known 
to date. Maxted et al. (\cite{maxted3}) reported the detection of a significant 
reflection effect, but no eclipses in the light curve. Taking these informations into 
account, one can constrain the inclination angle to be lower than $73^{\circ}$ 
(no eclipses!) and derive a minimum sdB mass of $0.47\,M_{\rm \odot}$. 
The minimum mass of the companion is constrained to $0.06\,M_{\rm \odot}$.
 The companion is therefore most likely a brown dwarf (BD) or a very late 
 M dwarf. Only two other candidate sdB+BD systems are known. 
 
{\bf HS\,0705$+$6700} is an eclipsing sdB+M binary with reflection effect. 
Drechsel et al. (\cite{drechsel}) performed a detailed photometric and 
spectroscopic analysis of this system and derived an inclination of $84^{\circ}.
4$, an sdB mass of $0.483\,M_{\rm \odot}$ and a companion mass of 
$0.134\,M_{\rm \odot}$. 
Drechsel et al. (\cite{drechsel}) also estimated $v_{\rm rot}\sin{i}$ and 
derived the companion mass. Although our result is much less accurate 
($0.15_{-0.03}^{+0.05}\,M_{\rm \odot}$), it comes close to that derived from the 
light curve. 

Much better agreement is reached for {\bf HW\,Vir}, the prototype eclipsing 
sdB+M binary, where excellent high resolution spectra are available. 
Edelmann (\cite{edelmann3}) recently determined the absolute parameters 
of this system spectroscopically using shallow absorption lines of the secondary 
to obtain its RV curve for the first time.\footnote{Wood \& Saffer (\cite{wood2}) detected these features in low resolution spectra before.} Edelmann (\cite{edelmann3}) derives 
an sdB mass of $0.53\,M_{\rm \odot}$ and a companion mass of 
$0.15\,M_{\rm \odot}$. Adopting this sdB mass our derivation of the companion 
mass agrees very well ($0.155_{-0.015}^{+0.015}\,M_{\rm \odot}$). The derived inclination 
angle of $i=75_{-10}^{+15}\,^{\circ}$ is consistent with the more accurate photometric 
solution $i=80^{\circ}.6\pm0^{\circ}.2$ given by Wood, Zang \& Robinson 
(\cite{wood}). Most recently Lee et al. (\cite{lee}) presented an analysis on HW\,Vir based on new photometric 
data. Their best solution ($i=80^{\circ}.98\pm0^{\circ}.1$, $M_{\rm 1}=0.485\pm0.013\,M_{\rm \odot}$, 
$M_{\rm 2}=0.142\pm0.004\,M_{\rm \odot}$) is fully consistent with our results.

The eclipsing and pulsating sdBV+M binary {\bf PG\,1336$-$018} (NY\,Vir) 
has been analysed by Vu\v ckovi\'c et al. (\cite{vuckovic}), but no unique 
solution could be found. In an asteroseismic study Charpinet et al.
 (\cite{charpinet5}) derived the fundamental parameters of this star by 
 fitting simultaneously the observed pulsation modes detectable in the 
 light curve. Adopting the asteroseismic value for the sdB mass
 ($0.459\,M_{\rm \odot}$) for our analysis, the companion mass is 
 $>0.12\,M_{\rm \odot}$. This result is in agreement with the second solution 
 from Vu\v ckovi\'c et al. 
 (\cite{vuckovic}): $M_{\rm sdB}=0.467\,M_{\rm \odot}$, 
 $M_{\rm comp}=0.122\,M_{\rm \odot}$. Charpinet et al. (\cite{charpinet5}) 
 concluded that the binary must be synchronised to account for the observed 
 rotational splitting of the pulsation modes and predict 
 a $v_{\rm rot}\sin{i}=74.9\pm0.6\,{\rm kms^{-1}}$. This predicted value is 
 consistent with the derived upper limit of 
 $v_{\rm rot}\sin{i}<79\,{\rm kms^{-1}}$. 

{\bf BPS\,CS\,22169$-$0001} was proposed to host a BD companion 
(Edelmann et al. \cite{edelmann}), but we derived a very low inclination and
 therefore a companion mass too high for a BD ($0.19_{-0.06}^{+0.07}\,M_{\rm \odot}$). 
In the light curves of the four binaries {\bf BPS\,CS\,22169$-$0001}, 
{\bf HE\,0230$-$4323}, {\bf JL\,82} as well as {\bf PG\,1329$+$159} reflection
 effects have been detected (see references in Table~\ref{compmasses}). 
 The derived companion mass ranges are consistent with the masses of late M 
 dwarfs. 

\subsection{White dwarfs \label{sec:lowmasswd}}

Ten stars must have white dwarf companions because no lines from cool companions are 
visible and the absence of a reflection effect can be used to exclude a
main sequence companion in some cases.

Among these binaries {\bf KPD\,1946$+$4340} sticks out. Most recently Bloemen et al. (\cite{bloemen}) discovered eclipses and ellipsoidal variations in a spectacular high precision light curve obtained by the Kepler mission. The eclipses are clearly caused by a WD  companion. We derive a mass range of $0.59-0.85\,M_{\rm \odot}$ for the unseen companion consistent with a WD. Due to the fact that the binary is eclipsing, the inclination angle has to be close to $90^{\rm \circ}$. Assuming the canonical sdB mass of $0.47\,M_{\rm \odot}$ the companion mass can be constrained to $\simeq0.61\,M_{\rm \odot}$, which is the average mass of WDs with C/O core. This  result is perfectly consistent with the indepedent analysis of Bloemen et al. (\cite{bloemen}). 

The companion of {\bf GD\,687} has already been shown to be a white dwarf
by Geier et al. (\cite{geier4}) utilising the same technique as used in this
paper and is included for the sake of completeness. 
Its merging time of $11.1\,{\rm Gyr}$,
 which is just a little shorter than the Hubble time.\footnote{The merging times of all binaries have been calculated using the formula given in Ergma et al. (\cite{ergma}).} 
 
A remarkable object which has a high inclination and a very low companion mass 
($>0.10\,M_{\rm \odot}$) is {\bf PG\,1043$+$760}. Due to its short period of 
$0.12\,{\rm d}$ a reflection effect should be easily detectable. 
But Maxted et al. (\cite{maxted5}) report a non-detection of variations 
in the light curve. The companion of this star must be a compact object, most 
likely a helium-core white dwarf of very low mass. 

\begin{figure}[]
\begin{center}
	\resizebox{\hsize}{!}{\includegraphics{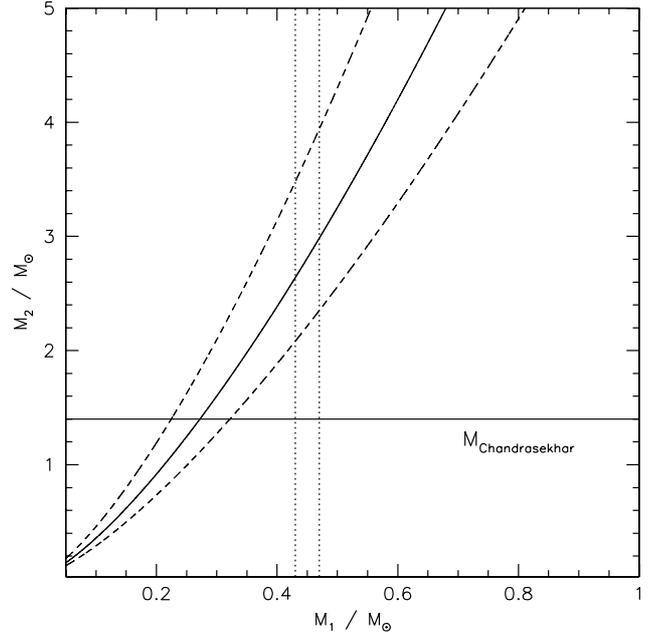}}
	\caption{Mass of the sdB primary HE\,0532$-$4503 plotted against the mass of the unseen companion. The companion mass error is indicated by the dashed lines. The mass range of the CE ejection channel (Han et al. \cite{han1}) is marked with dotted vertical lines. The Chandrasekhar mass limit is plotted as solid horizontal line.}
	\label{he0532_mass}
\end{center}
\end{figure}

In the case of {\bf PG\,1627$+$017}, a main sequence companion can be excluded 
as well. With a mass exceeding $0.50\,{M_{\rm \odot}}$ the companion would be 
visible in the spectra in this case. The non-detection of a reflection effect 
(Maxted et al. \cite{maxted5}; For et al. \cite{for}) is consistent with our result.

The companion of {\bf PG\,0101$+$039} is a white dwarf. Despite of the long
 orbital period of $0.57\,{\rm d}$ a main-sequence companion could be excluded. 
 A light curve was taken with the MOST satellite. Instead of a reflection 
 effect the shallowest ellipsoidal deformation ever detected could be 
 verified (Geier et al. \cite{geier2}). The white dwarf companion could be 
 quite massive ($0.52-0.92\,{M_{\rm \odot}}$). In this case the total mass 
 comes close the Chandrasekhar limit, but the merging time would be higher 
 than the Hubble time. PG\,0101$-$039 does therefore not qualify as SN\,Ia 
 progenitor candidate. The companion mass range of {\bf PG\,0001$+$275} is 
 quite similar ($0.56-1.05\,{M_{\rm \odot}}$). A main sequence companion 
 can be most likely excluded and no reflection effect was detected (Maxted et al. \cite{maxted5}; Shimanskii et al. \cite{shimanskii}). The orbital period of $0.53\,{\rm d}$ is also too long to make PG\,0001$+$275 an SN\,Ia progenitor candidate.

\begin{figure}[t!]
\begin{center}
	\resizebox{\hsize}{!}{\includegraphics{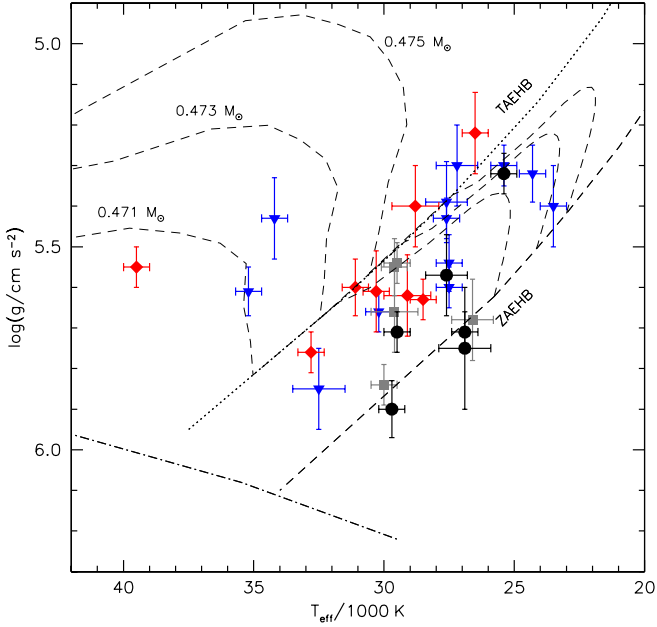}}
	\caption{$T_{\rm eff}-\log{g}$-diagram; same as Fig. \ref{fig:tefflogg}
	but restricted to the sample which could be solved under the assumption of synchronisation. The helium main sequence and EHB band are superimposed with EHB evolutionary tracks from Dorman et al. (\cite{dorman}) labelled with their masses. Binaries with confirmed late main sequence or brown dwarf companions are plotted as filled diamonds, binaries with confirmed white dwarf companions with filled triangles. 
	Hot subdwarfs where the companion could be a main sequence star or a white 
	dwarf are marked with solid rectangles. The filled circles mark the sdBs 
	with putative massive compact companions. 
	}
	\label{fig:teffloggsync}
\end{center}
\end{figure}

\begin{figure}[t!]
\begin{center}
	\resizebox{\hsize}{!}{\includegraphics{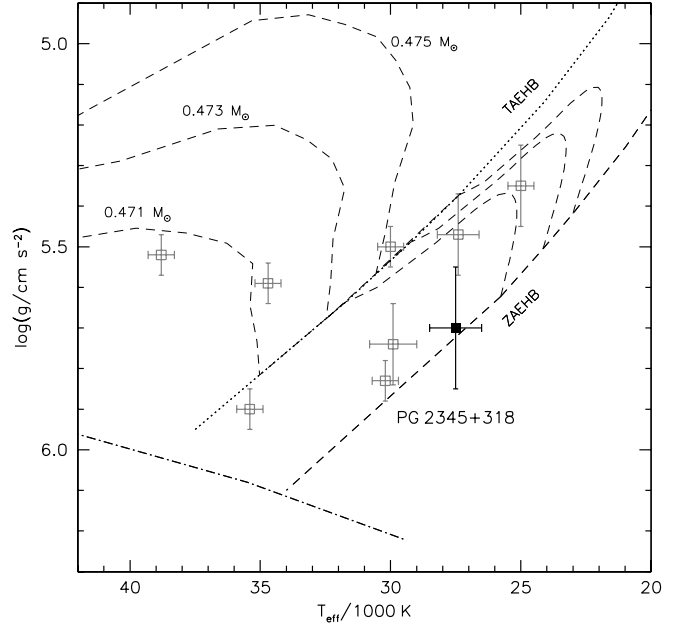}}
	\caption{$T_{\rm eff}-\log{g}$-diagram; same as Fig. \ref{fig:teffloggsync}
	but restricted to the non-synchronised systems. The open squares mark binaries with orbital periods longer than $1.2\,{\rm d}$. The filled one marks the system where synchronisation is not established despite its short orbital period.}
	\label{fig:teffloggnosync}
\end{center}
\end{figure}

Edelmann et al. (\cite{edelmann}) derived a very low minimum companion mass 
for {\bf CPD\,$-$64\,481}. At high inclination the companion mass would have 
been consistent with a brown dwarf. However, our analysis provides evidence 
that this binary has a very low inclination ($i=5^\circ$ to $9^\circ$), actually
the lowest one of the entire sample,  
and therefore a companion mass way too 
high for a BD ($0.62_{-0.24}^{+0.42}\,M_{\rm \odot}$) indicating a white dwarf binary. 
Due to the low projected rotational 
velocity of this star, the fractional error is very high and the companion 
mass not very well constrained. For the highest possible companion mass the 
system would exceed the Chandrasekhar limit and qualify as SN\,Ia progenitor 
candidate due to its short orbital period. However, the inclination angle must 
be lower than $5^{\circ}$ is this case. That is why this extreme scenario is 
considered to be very unlikely.

The unseen companions in the binaries {\bf HE\,1047$-$0436}, 
and {\bf HD\,171858} also have masses consistent with white dwarfs.

\begin{figure*}[t!]
\begin{center}
	\resizebox{\hsize}{!}{\includegraphics{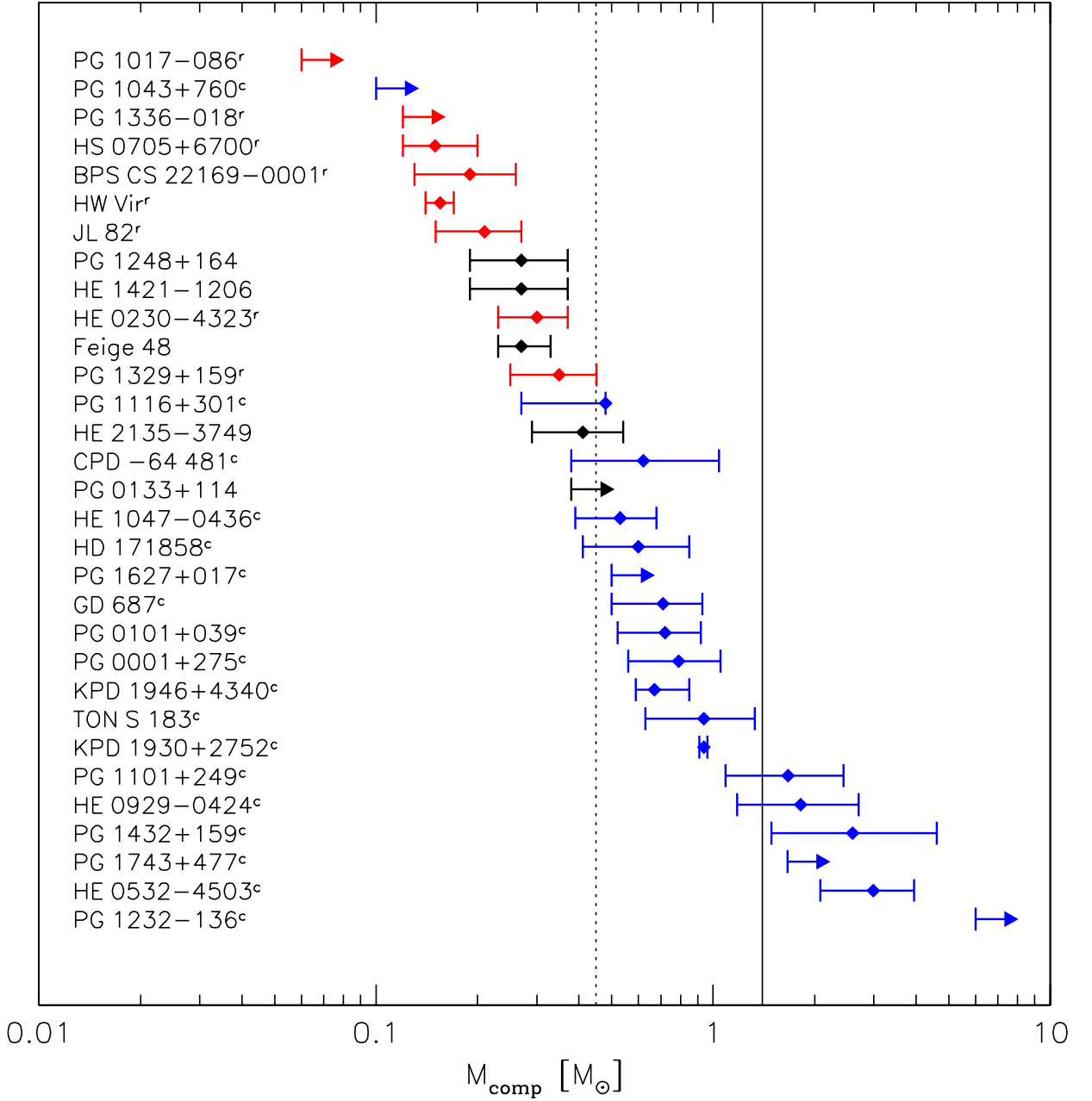}}
	\caption{Mass ranges for the unseen companions of 31 binaries under 
	the assumption of synchronisation (see Table~\ref{compmasses}). The 
	companion mass ranges are derived for the most likely sdB mass range 
	of $0.43-0.47\,M_{\rm \odot}$. The dashed vertical line marks the 
	upper limit to the mass of main-sequence companions. Main-sequence stars with higher masses 
	would be visible in the spectra and can be excluded. The solid vertical
	 lines marks the Chandrasekhar mass limit. $^{r}$Binaries with 
	 reflection effect detected in their light curves. The companions are
	  either late M stars or brown dwarfs. $^{c}$Binaries with compact 
	  companions like white dwarfs, neutron stars or black holes.}
	\label{figcompmasses}
\end{center}
\end{figure*}

The mass of the companion to {\bf PG\,1116$+$301} is slightly above the limit of 
$0.45\,{M_{\rm \odot}}$. Despite the high inclination derived for this binary no reflection effect was detected in its light curve (Maxted et al. \cite{maxted5}; Shimanskii et al. \cite{shimanskii}), which is consistent with a WD companion.\footnote{The upper limit to the companion mass of PG\,1116$+$301 is identical with the most likely companion mass (see Table~\ref{compmasses}, Fig.~\ref{figcompmasses}). The system can only be synchronised if the inclination reaches its maximum value of $90^{\rm \circ}$. In this case the upper limit to the sdB mass is lower than $0.47\,M_{\rm \odot}$, but still within the possible range (see Sect.~\ref{sec:ana}).}

\subsection{Massive compact companions - white dwarfs, neutron stars, black 
holes \label{sec:highmassbh}}

Seven subdwarf binaries (in addition to KPD~1930+2752) have 
massive compact companions (see e.g. Fig. \ref{he0532_mass}) exceeding 
$0.9\,M_{\rm \odot}$. For all of 
these binaries main sequence companions can be excluded, because they would 
significantly contribute to the flux or even outshine the subdwarf primary. 
The massive companions therefore have to be compact. 

The nature of the unseen companion in the binary {\bf KPD\,1930$+$2752} could 
be clarified by Geier et al. (\cite{geier}). The short period system consists 
of a synchronously rotating, tidally distorted sdB and a massive white dwarf. The 
combined mass of the systems reaches the Chandrasekhar limit and the stars 
will most probably merge in $200\,{\rm Myr}$. KPD\,1930$+$2752 is the best 
double degenerate candidate for SN\,Ia progenitor so far. 

The companion mass of {\bf TON\,S\,183} is as high as that of KPD\,1930$+$2752.
However, the error bar is much larger. Hence we can not exclude that it is a
normal white dwarf of $0.6\,M_{\rm \odot}$. 
On the other hand the total mass of the system may exceed the 
Chandrasekhar limit, but TON\,S\,183 does also not qualify as SN\,Ia progenitor 
candidate, because of its long orbital period the merging time exceeds the 
Hubble time by orders of magnitude.

For {\bf PG\,1101$+$249} and  {\bf HE\,0929$-$0424} the companion mass is slightly above the Chandrasekhar limit, but we can not exclude a massive white dwarf given the errors. 
 The merging times of HE\,0929$-$0424 and especially PG\,1101$+$249 on the other hand would be near 
or below Hubble time and the total masses of the systems would most likely 
exceed the Chandrasekhar limit. If the companions should be massive white 
dwarfs of C/O composition, these binaries would be SN\,Ia progenitor candidates. 

The companions of {\bf PG\,1432$+$159}, {\bf HE\,0532$-$4503} and 
{\bf PG\,1743$+$477} may be neutron stars as well as black holes as their 
masses exceed the Chandrasekhar limit even when errors are accounted for. Light curves have been obtained of both  PG\,1432$+$159 and PG\,1743$+$477. The non-detection of reflection effects is perfectly consistent with compact companions (Maxted et al. \cite{maxted5}). In the case of  PG\,1743$+$477 only a lower limit for the companion mass could be derived. Due to their short orbital periods the companions in PG\,1432$+$159 as well as in HE\,0532$-$4503 will merge in a few billion years at most. Since the average 
lifetime on the EHB is only $100\,{\rm Myr}$ the sdBs will evolve to white 
dwarfs in the meantime. The outcome of a merger between a white dwarf and a neutron star 
or a black hole is unclear. Such systems may be progenitors for gamma-ray 
bursts or more exotic astrophysical transients (see discussion in Badenes et 
al. \cite{badenes}).

In the case of {\bf PG\,1232$-$136} only a lower limit can be given for the companion mass 
($>6.0\,M_{\rm \odot}$) which is higher than all theoretical NS masses. The 
companion of this sdBs may therefore be a BH. 

\subsection{Distribution in the $T_{\rm eff}$-$\log{g}$-plane \label{sec:distribtefflogg}}

Fig.~\ref{fig:teffloggsync} shows the distribution of the 31 solved binaries in the $T_{\rm eff}$-$\log{g}$-diagram. Within their error bars most of the sdB primaries are associated with the EHB as expected. Only three of them  (BPS\,CS\,22169$-$0001, KPD\,1930$+$2752, KPD\,1946$+$4340) have evolved beyond the TAEHB. No trends with companion types can be seen. The location on the EHB is a function of the thickness of the stars' hydrogen layers. The thinner this layer is, the higher are $T_{\rm eff}$ and $\log{g}$ at the beginning of EHB-evolution and the more envelope mass has been lost during the CE-ejection. The efficiency of this process seems to be not much affected by the companion type. Companions of all types ranging from low mass M dwarfs or brown dwarfs to massive compact objects are scattered all over the EHB. 

While the fraction of evolved sdBs is only $10\%$ in the solved sample, two out of nine subdwarfs ($22\%$) are found in binaries, which could not be solved under the assumption of synchronisation, are obviously not located on the EHB (see Fig.~\ref{fig:teffloggnosync}). A possible reason for this discrepancy is discussed in Sect.~\ref{sec:hd188}.

\subsection{Distribution of companion masses \label{sec:distribsystem}} 

\begin{figure}[t!]
\begin{center}
	\resizebox{\hsize}{!}{\includegraphics{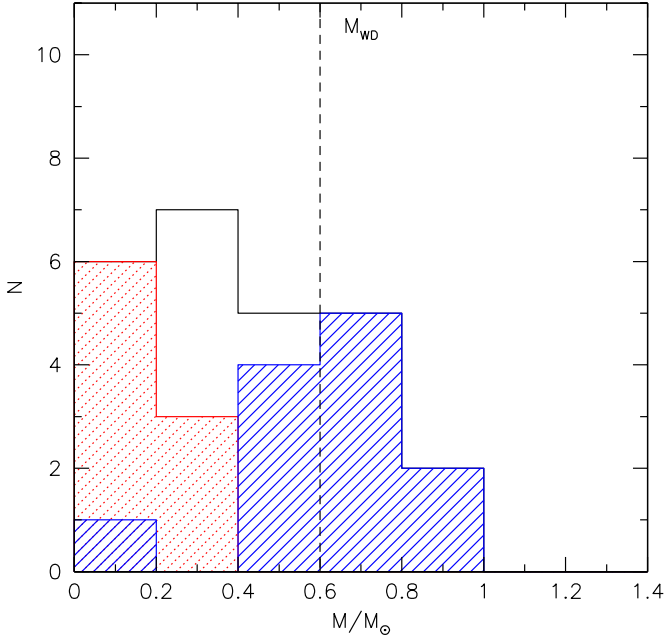}}
	\caption{Companion mass distribution of the binaries with low mass 
	companions (Table~\ref{compmasses}, 
	Fig.~\ref{figcompmasses}). 
	The solid histogram shows the fraction of 
	subdwarfs with confirmed white dwarf companions, the dashed histogram the 
	detected M dwarf companions. The dashed vertical line marks the 
	average white dwarf mass.}
	\label{massdistribdetail}
\end{center}
\end{figure}

Fig.~\ref{massdistribdetail} shows the low mass end of the companion mass distribution. 
Excluding the massive systems described in Sect.~\ref{sec:highmassbh} 
the histogram mass distribution (Fig.~\ref{massdistribdetail}) displays a peak 
at companion masses ranging from $0.2-0.4\,M_{\rm \odot}$. 
Most of the low mass objects $<0.4\,M_{\rm \odot}$ have been identified as M dwarfs. 
The bona fide white dwarf companions seem to peak at masses ranging from $0.4\,M_{\rm \odot}$ to $0.8\,M_{\rm \odot}$. Because close binary evolution is involved, there should be deviations from the normal mass distribution of single white dwarfs, which shows a characteristic peak at an average mass of $0.6\,M_{\rm \odot}$. We therefore conclude that the mass distribution of the restricted sample looks reasonable and no obvious systematics can be seen. The high fraction of massive compact companions (up to $20\%$ of our sample) on the other hand looks suspicious. 

As the companion mass depends on the primary mass, the companion masses would be
lower, if the primaries' masses were overestimated.
We have adopted the masses of the sdB primaries to range from $0.43\,M_{\rm \odot}$
to $0.47\,M_{\rm \odot}$ as suggested by the models of Han et al. 
(\cite{han1}, \cite{han2}) and backed-up by asteroseismology. 
However, the minimum mass of a core helium burning 
star can be as small as $0.3\,M_{\rm \odot}$. 

\begin{figure}[t!]
\begin{center}
	\resizebox{\hsize}{!}{\includegraphics{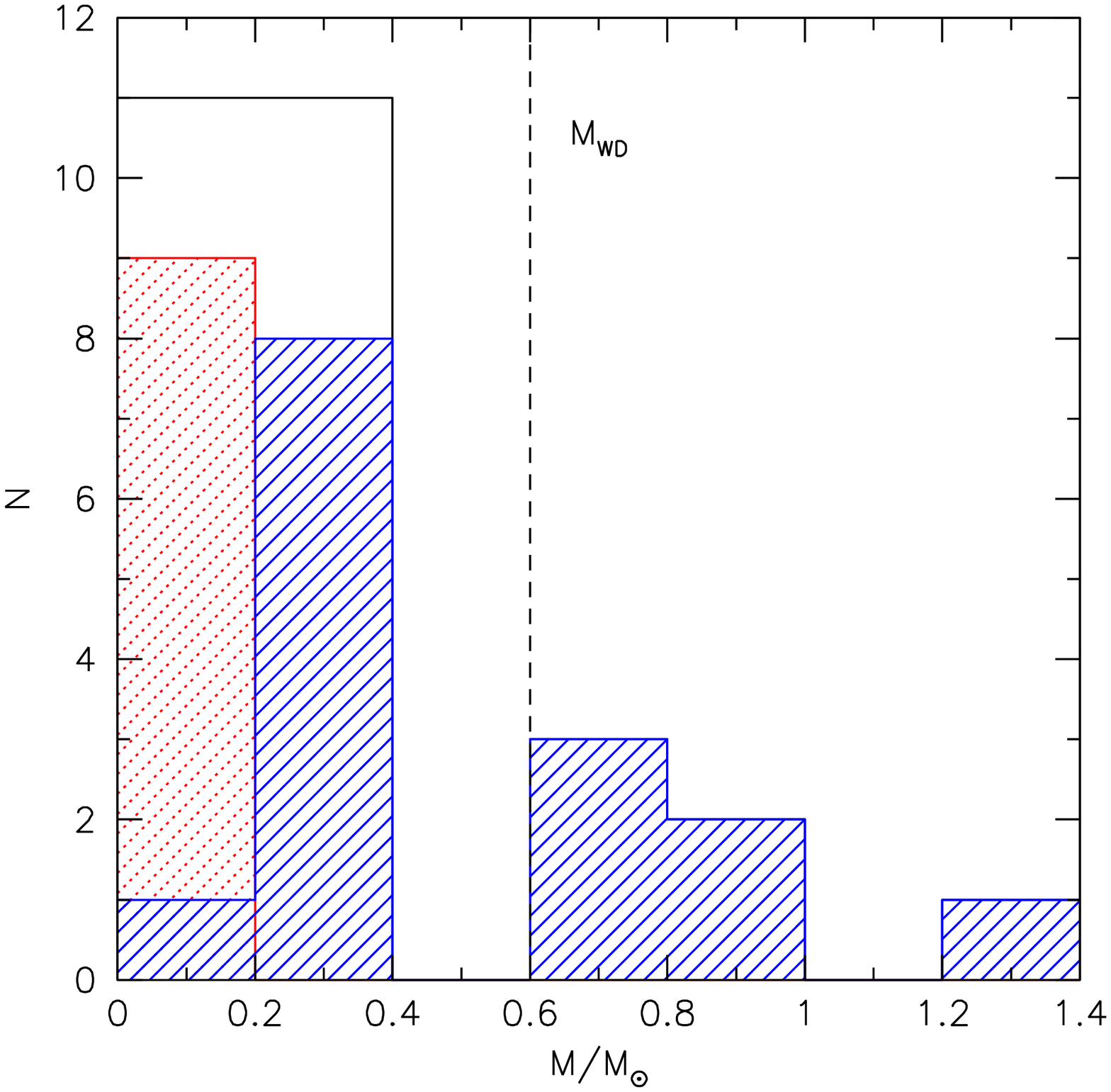}}
	\caption{Mass distribution of the unseen companion stars 
	(see Fig.~\ref{massdistribdetail}). 
	The lowest possible companion mass is plotted against the total 
	number of binaries under the assumption of the lowest possible sdB mass.}
	\label{massdistrib030}
\end{center}
\end{figure}

In Fig.~\ref{massdistrib030} the companion mass distribution is plotted under
the extreme assumption that all sdBs have this minimum mass for core helium 
 burning (or the minimum mass allowed by other constraints). Looking at the 
 low mass regime and comparing the distribution with 
  Fig.~\ref{massdistribdetail} one immediately notices that this assumption 
  leads to unphysical results. The distribution of low mass companions peaks 
  at masses lower than $0.4\,M_{\rm \odot}$, which is very unlikely 
  especially for white dwarf companions. 

Under this extreme assumption only the companion of PG\,1232$-$136 remains 
more massive than the Chandrasekhar limit. 
Furthermore the companions of PG\,1743$+$477 and HE\,0532$-$4503 still are 
more massive than $1.0\,M_{\rm \odot}$ in this case. With just slightly 
higher sdB masses the companion masses would exceed the Chandrasekhar limit.

\subsection{The inclination problem}  

\begin{figure}[t!]
\begin{center}
	\resizebox{\hsize}{!}{\includegraphics{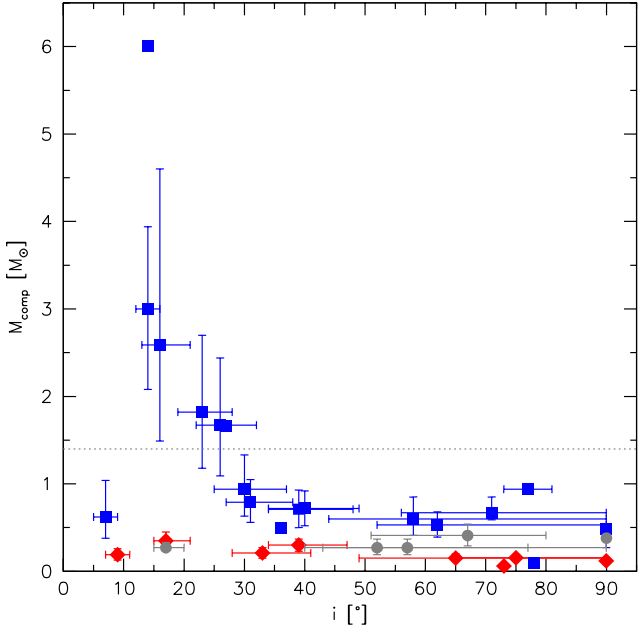}}
	\caption{Companion mass versus inclination. The solid squares mark compact companions (WD/NS/BH), the solid diamond MS or BD companions. The solid circles mark objects where both companion types are possible.}
	\label{massincl}
\end{center}
\end{figure}

By plotting the companion masses versus inclination angles 
(Fig.~\ref{massincl}) an anomaly becomes apparent. While the systems with 
low mass companions cover all inclination angles with a slight 
preference for high inclinations, the systems with massive compact companions
are found at low inclinations between $15^{\rm \circ}$ and $30^{\rm \circ}$.

Our sample has been drawn from the catalogue of Ritter \& Kolb (\cite{ritter}),
which is a compilation extracted from literature and not a systematic survey.
Hence selection effects can not be quantified. Most of the low-mass,
high-inclination systems have been discovered by photometry 
(eclipses and reflection effect), while all others stem from radial velocity
surveys. The radial velocity technique is biased against low inclinations and 
low masses. Hence, massive systems at high inclinations should be found most
easily. However, except for KPD\,1930$+$2752, there is no high inclination object
among the subsample of massive compact companions. 
One may speculate that such systems may have been overlooked,
because their spectra may look peculiar due to orbital smearing and are
therefore not classified as sdB stars. 
 
We refrain from further speculations about selection effects and proceed to 
search for an evolutionary scenario that can explain the formation of sdB 
binaries with neutron star or black hole companions.

\section{The formation of sdB+NS/BH binaries} \label{sec:form}

Neutron stars and stellar-mass black holes are the remnants of massive stars 
ending their lifes in supernova explosions. Detecting these exotic objects is 
possible when they are in a close orbit with another star. If matter is 
transferred from the companion star to the compact object, X-rays are emitted. 
Not many neutron stars or stellar mass black holes could be found up to now. 
On the other hand evolved, non-interacting binaries containing such objects 
should exist, since X-ray binaries only represent a relatively short phase of 
stellar evolution. Without ongoing mass transfer the companion remains 
invisible, but should be detectable indirectly from the reflex motion of the 
visible star. Badenes et al. (\cite{badenes}) discovered a 
massive compact companion to a white dwarf and concluded that this companion is likely to be a neutron star. 
But Marsh et al. (\cite{marsh}) convincingly showed that the system is a double degenerate system consisting of a low mass and a very high mass WD. Kulkarni \& van Kerkwijk (\cite{kulkarni}) performed an independent analysis with similar results. In this section the question whether sdB stars with hidden neutron star or black hole companions do exist is discussed in 
detail. 

\begin{figure*}[t!]
\begin{center}
\includegraphics[angle=90,width=18cm]{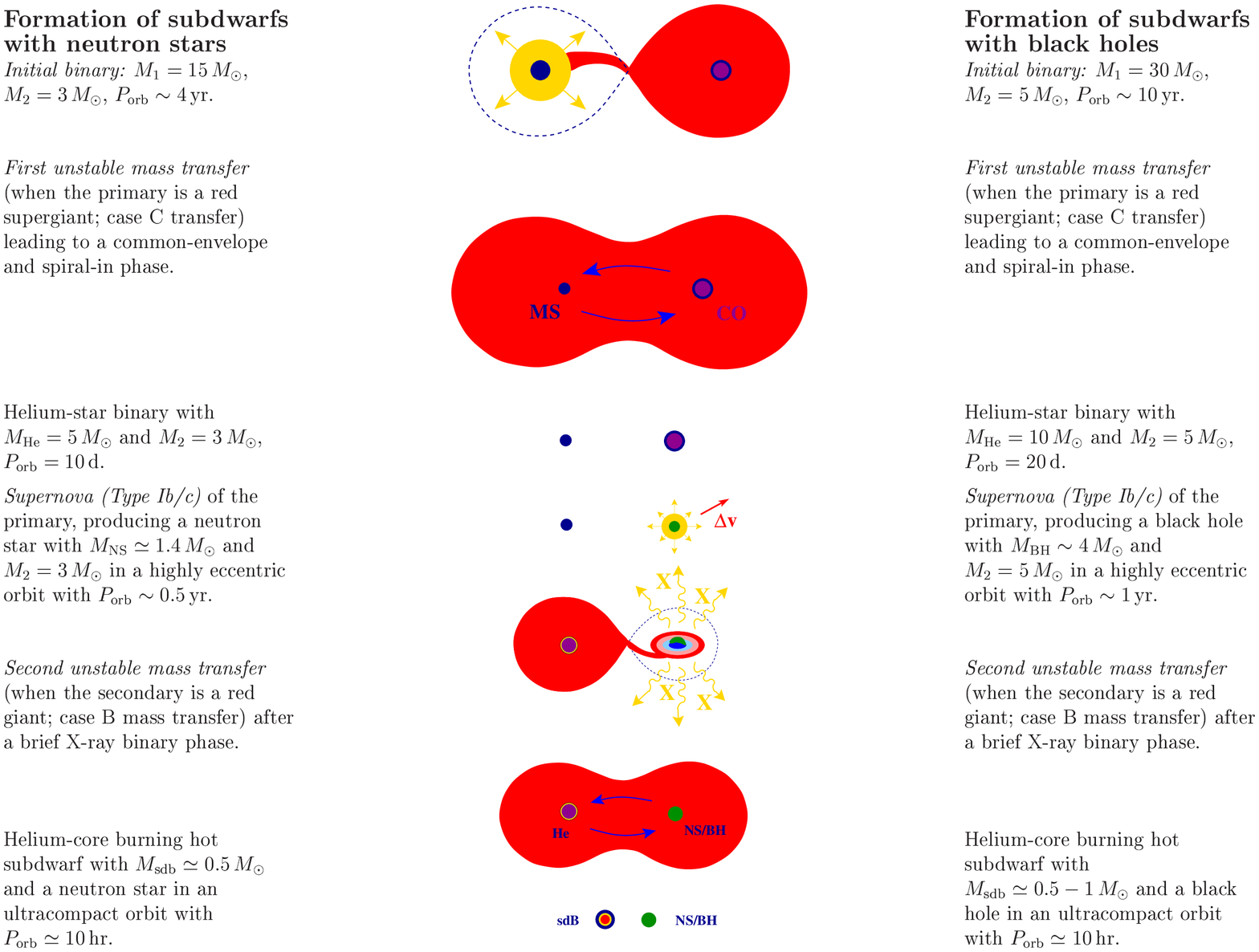}
 \caption{Schematic diagram of formation scenarios leading to hot
 subdwarf binaries with neutron-star (left hand panel) or black-hole (right
 hand panel) companions.}
 \label{formation}
\end{center}
\end{figure*}

The existence of sdB+NS/BH systems requires an appropriate formation channel. 
The evolution that leads to such systems requires an initial binary,
consisting of a primary star that is sufficiently massive to produce a
neutron star or black hole, and a companion, the progenitor of the hot
subdwarf, of typically several solar masses. The initial orbital
period has to be quite large (a few to 20 years), so that mass
transfer only starts late in the evolution of the star, and these
systems generally experience two mass-transfer phases and one
supernova explosion (see Fig.~\ref{formation}). The short orbital periods 
observed
for our systems imply that the second mass-transfer phase from the red
giant progenitor of the subdwarf to the compact companion had to be
unstable, leading to a common-envelope and spiral-in phase of the
compact object. The condition for unstable mass transfer constrains
the mass of the progenitor to be larger than the mass of the compact
object (otherwise, mass transfer would be stable and lead to a much
wider system, Podsiadlowski et al. \cite{podsi}). 
Fig.~\ref{formation} illustrates the evolution that leads to
systems of this type for two typical examples. While this scenario can
explain most of our systems with high-mass compact components, the
inferred masses of the putative black hole in PG\,1232$-$136 is  
larger than we would estimate ($\le 3\,M_\odot$) for a 0.5\,$M_\odot$
sdB star. This may suggest that this system has experienced another 
mass-transfer phase after the two common-envelope phases in which mass 
was transferred from the sdB star to the compact object. It should
also be noted that, while we assume here that
the mass of the subdwarf is $\sim 0.5\,M_\odot$, consistent with the
properties of the observed systems, the sdB mass range allowed by this
scenario is $0.3 - 1.1\,M_\odot$ for the neutron-star systems and $0.5
- 1.1\, M_\odot$ for the black-hole systems. Compared with the mass
range of $0.3 - 0.7\,M_\odot$ for the standard evolutionary
channel (Han et al. \cite{han1}, \cite{han2}), the subdwarf may therefore 
be more massive. An independent determination of the sdB mass (e.g. by 
obtaining parallaxes) could therefore help to verify this
scenario.

At the beginning of the second mass-transfer phase, these systems are
expected to pass through a short X-ray binary phase, lasting $\sim
10^5\,$yr, in which a neutron star may accrete up to $\sim
10^{-3}\,M_\odot$ and become a moderately recycled millisecond
pulsar (Podsiadlowski et al. \cite{podsi}). This links these system to the
X-ray binary population
(in a sense, they are failed low-mass X-ray binaries). Population
synthesis estimates (Pfahl et al. \cite{pfahl}) suggest that up to one in 
$10^4$ stars in
the Galaxy experience this evolution, implying that of order $1\,\%$ of
all hot subdwarfs should have neutron-star or black-hole
companions. This means that tens of thousands of these systems could
exist in the Galaxy compared to just about $300$ known X-ray binaries. 
The binary PSR\,J1802$-$2124, which consists of a millisecond pulsar and a CO white dwarf in close orbit ($P=0.7\,{\rm d}$, $M_{\rm WD}=0.78\,M_{\rm \odot}$) may have evolved in a similar way (Ferdman et al. \cite{ferdman}).

\section*{Part II: Synchronisation -- Theory and empirical evidence}

\section{Orbital synchronisation of sdB binaries \label{sec:tidal}}

The results presented above are based on the assumption of tidal 
synchronisation. Since especially the discovery of sdB+NS/BH systems 
challenges our understanding of stellar evolution, it is necessary to 
investigate whether this assumption holds in the case of sdBs. A thorough 
discussion of tidal synchronisation in sdB binaries both from the theoretical 
and the observational point of view is therefore given.

\subsection{Theoretical timescales for synchronisation}

Which mechanism is responsible for orbital synchronisation in binaries is 
still under debate. Theoretical timescales for synchronisation are given by
 Zahn (\cite{zahn2}) and Tassoul \& Tassoul (\cite{tassoul}), but 
 unfortunately they are not consistent for stars with radiative envelopes and 
 convective cores like hot subdwarfs.

Zahn (\cite{zahn2}) was the first to calculate synchronisation and 
circularisation timescales for main sequence stars in close binary systems. 
Observations of eclipsing binaries were in good agreement with his theoretical 
calculations for late type main-sequence stars with radiative cores and convective 
envelopes. Tidal friction caused by the equilibrium tide, which forms under 
the tidal influence of the close companion, is very efficient in this case 
because convection connects the inner regions of the stellar envelope with its 
surface. For radiative envelopes another mechanism is needed to explain the 
observed degree of synchronism in early type main-sequence binaries. Dynamical tides, 
which are excited at the boundary layer between the convective core and the 
radiative envelope are thought to be radiatively damped at the stellar surface 
and to transfer angular momentum outwards. This mechanism turns out to be much 
less efficient and the predicted synchronisation timescales are too long to 
explain the degree of synchronism in some early type main-sequence stars (e.g. Giuricin 
et al. \cite{giuricin}). 

Tassoul \& Tassoul (\cite{tassoul}) introduced another, hydrodynamical braking 
mechanism. Tidally induced meridional currents in the non-synchronous binary 
components should lead to synchronisation and circularisation of the system. 
This mechanism is very efficient, but it was debated whether it is valid or 
not (Rieutord \cite{rieutord}; Tassoul \& Tassoul \cite{tassoul2}). Claret et 
al. (\cite{claret1}, \cite{claret2}) studied both mechanisms and compared them 
to the available observations. Due to the necessary calibration of many 
uncertain parameters a definitive answer as to which mechanism is in better 
agreement with observation could not be given. 

Applying the theory of tidal synchronisation to sdB binaries is not an easy 
task. One of the key results of both theories is that tidal circularisation of 
the orbit is achieved after the companions are synchronised. 
This means that 
once an orbital solution is found and the orbit turns out to be circular, both 
companions can be considered as synchronised without knowing their rotational 
properties. This simple law cannot be used in the case of sdBs. The reason is
that close binary sdBs were formed via the CE ejection channel. 
The common envelope phase is very efficient in circularising the orbit and 
all known close binary sdBs have circular orbits or show only small  
eccentricities ($\epsilon \leq 0.06$; Edelmann et al. \cite{edelmann}; 
M\"uller et al. \cite{mueller}; Napiwotzki et al. in prep.). 

Stellar structure plays an important role. The synchronisation timescale of 
Zahn (\cite{zahn2}) scales with $(R_{\rm C}/R)^{8}$, where $R_{\rm C}$ is the 
radius of the convective core and $R$ the stellar radius. The larger the
convective core of a star, the shorter the time span until synchronisation is 
reached. 

In order to estimate the synchronisation times of the analysed binaries we 
used the formulas of Zahn (\cite{zahn2}) and Tassoul \& Tassoul 
(\cite{tassoul}). 

\begin{eqnarray}
\label{eq:zahn}
t_{\rm sync}({\rm Zahn})&=&52^{-5/3}\left(\frac{R^{3}}{GM}\right)^{1/2}
             \left(\frac{I}{MR^{2}}\right)\nonumber\\
             & &\times\frac{\left(1+q\right)^{5/6}}{q^{2}}
             E_{2}^{-1}\left(\frac{a}{R}\right)^{17/2}
\end{eqnarray}

Here $M=M_{\rm sdB}$, $R=R_{\rm sdB}$, $q=M_{\rm comp}/M_{\rm sdB}$, $a$ is 
the separation of the companions, which can be calculated from the measured 
orbital parameters using Kepler's third law, and $I$ is the moment of inertia 
of the sdB star. We adopted the canonical sdB mass 
($M_{\rm sdB}=0.47\,M_{\rm \odot}$) for these calculations. $E_{\rm 2}$ is a
 tidal coefficient which is very sensitive to the structure of the star, 
 especially the size of the convective core. Here we use the first 
 approximation of Zahn (\cite{zahn2}) $E_{\rm n}=(R_{\rm C}/R)^{2n+4}$ and 
 adopt $R_{\rm C}/R \simeq 0.15$ and $\frac{I}{MR^{2}}\simeq 0.04$ derived from sdB models calculated by Han 
 (priv. comm.). For this models a hydrogen layer mass of $10^{-4}\,M_{\rm \odot}$ was chosen 
 consistent with result from asteroseismology (e.g. Charpinet et al.
  \cite{charpinet5}).

\begin{eqnarray}
\label{eq:tassoul}
t_{\rm sync}({\rm Tassoul})&=&5.35\times10^{2+\gamma-N/4}
             \frac{1+q}{q}L^{-1/4}\nonumber\\
             & &\times M^{5/4}R^{-3}P^{11/4}
\end{eqnarray}

In this equation $M$, $R$ (solar units) and $q$ are defined in the same way as
 above. $P$ is the orbital period in days. The luminosity 
 $L=4\pi\sigma R^{2}T_{\rm eff}^{4}$ can be calculated using the 
 $T_{\rm eff}$ measurements given in Table~\ref{orbit}. The parameter $N$ is 
 connected with the different ways of energy transport within the outer layers 
 of the stellar envelope. It is assumed to be zero in stars with radiative 
 envelopes. The parameter $\gamma$ can be adjusted to account for large 
 deviations from synchronism and contributions of both companions. Here the 
 value $\gamma=1.6$ used by Claret et al. (\cite{claret1}) was chosen. 
It has to be noted that this approach is only a crude approximation. 
As stated by Claret et al. (\cite{claret2}), the differential equations which 
govern the orbital parameters of a binary must be integrated. For this EHB 
evolution has to be taken into account. A detailed study of this problem is 
beyond the scope of this paper and we shall use equations \ref{eq:zahn} and \ref{eq:tassoul} to 
estimate the timescale of synchronisation.

It has to be pointed out that both theories predict the synchronisation timescale 
to increase strongly with increasing orbital period and to 
decrease with increasing sdB radius as 
$t_{\rm sync}\sim P^\alpha$ and $t_{\rm sync}\sim R^{-\beta}$. In the 
theory of Zahn (\cite{zahn2}) the exponents are $\alpha=17/3$ and $\beta=9$,
while the Tassoul \& Tassoul formula gives $\alpha=11/4$ and $\beta=3$.
In addition the synchronisation timescale decreases as the mass ratio 
increases. Hence it will take lower mass companions longer to synchronise the sdB
star if the other parameters are constant. 

\subsection{Synchronisation of our sample}\label{sec:syncsample}

\begin{figure}[t!]
\begin{center}
	\resizebox{\hsize}{!}{\includegraphics{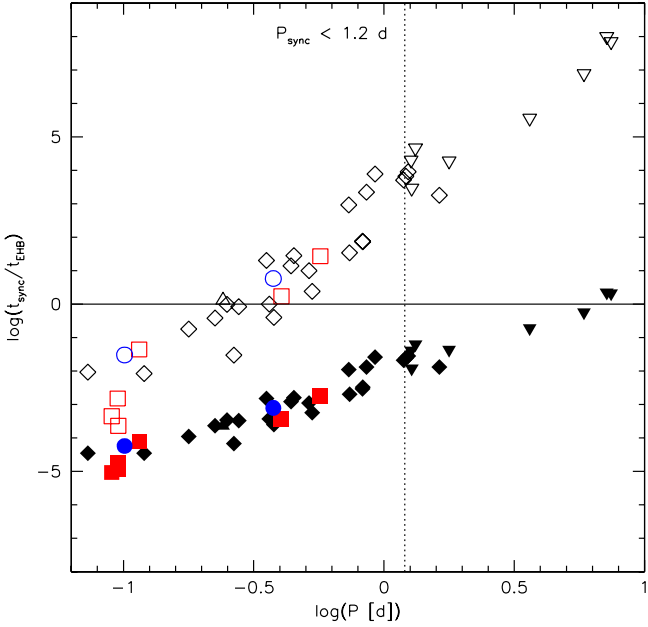}}
	\caption{Observed orbital period is plotted against the 
	synchronisation times of Zahn (\cite{zahn2}, open symbols) and 
	Tassoul \& Tassoul (\cite{tassoul}, filled symbols) both in units of 
	the average lifetime on the EHB ($10^{8}\,{\rm yr}$, Dorman et al. 
	\cite{dorman}). The solid horizontal line marks the border between 
	synchronisation within the EHB lifetime and synchronisation times 
	longer than the EHB lifetime. The squares mark sdB binaries, where the 
	primaries have been proven to be synchronised by light curve analysis 
	of eclipsing or ellipsoidal variable systems. The circles mark 
	binaries where synchronisation could be shown by asteroseismology. The 
	systems marked with diamonds could be solved consistently under the 
	assumption of synchronisation, while the systems marked with downward triangles 
	rotate faster than synchronised. PG\,2345$+$318 is the only sdB in our sample that rotates slower 
        than synchronised. It is marked with an upward triangle.}
	\label{fig:tsync}
\end{center}
\end{figure}

The synchronisation time scale depends strongly on orbital period and 
radius. Because the radii of the sdBs differ only little  
we display the results of our calculations as a function of orbital period in 
Fig. \ref{fig:tsync}.  
The synchronisation time scales  are given in units of the average EHB 
lifetime ($t_{\rm EHB}\simeq 10^{8}\,{\rm yr}$; Dorman et al. \cite{dorman}). 
A binary is thought to be synchronised, if the EHB 
lifetime is much longer than the synchronisation time. Due to the larger exponents $\alpha$ and $\beta$ the slope of the 
relations is steeper and the scatter larger for the Zahn (\cite{zahn2}) theory compared 
to the one proposed by Tassoul \& Tassoul (\cite{tassoul}). 
 What can be seen immediately is that the timescales of Zahn (\cite{zahn2}) and Tassoul \& Tassoul (\cite{tassoul}) 
differ by $2-8$ orders of magnitude.  
Observational evidence is needed to constrain the timescales of tidal 
synchronisation in close binary sdBs.

For periods shorter than $\simeq0.3-0.4\,{\rm d}$ both theories predict synchronised 
rotation and are consistent with our observations. In the period range 
$0.4-1.2\,{\rm d}$ only the synchronisation times of Tassoul are consistent 
with observation, while the timescales of Zahn quickly exceed Hubble time. 
If the orbital periods exceed $\simeq1.2-1.6\,{\rm d}$ the assumption of 
synchronisation does not yield consistent results any more, although the 
timescales calculated with the prescription of Tassoul \& Tassoul 
(\cite{tassoul}) would still predict synchronised rotation. 

According to our results, the period limit where synchronisation breaks down, 
lies near $1.2\,{\rm d}$. The binaries HE\,2150$-$0238 ($P=1.32\,{\rm d}$) 
and PG\,1512$+$244 ($P=1.2\,{\rm d}$) cannot be solved consistently although their periods 
are only slightly longer than that of HE\,1047$-$0436 ($1.21\,{\rm d}$) and 
PG\,0133$+$114 ($1.24\,{\rm d}$), which can be solved. 

Despite its long period, HD\,171858 can be solved 
consistently, making it the longest period ($P=1.6\,{\rm d}$) object in our sample 
that is synchronised. Why is this? Besides the orbital period the size of the 
star matters: The larger the star, the shorter the synchronisation time (see equations \ref{eq:zahn} and \ref{eq:tassoul}). 
The gravity of HD\,171858 is lower than that of all other stars with periods
ranging from $1.2\,{\rm d}$ to $1.6\,{\rm d}$ by a factor of $2$ at least. Hence its radius is larger 
and synchronisation can be achieved more quickly than in the other stars of slightly shorter periods. 

[CW\,83]\,1735$+$22 stands out among the longer-period binaries, because its
projected rotational velocity ($44\,{\rm km\,s^{-1}}$) is unusually high. 
Because of its period ($P=1.28\,{\rm d}$) it is not necessarily expected to be synchronised. 
This system is discussed in detail in Sect.~\ref{sec:hd188}.

We also found that the short period binary PG\,2345$+$318 ($P=0.24\,{\rm d}$) rotates slower than synchronised. 
This peculiar system is discussed in detail in Sect.~\ref{sec:age}.

In general the synchronisation mechanism of Zahn (\cite{zahn2}) is not 
efficient enough to explain the observed level of synchronisation, while the 
mechanism of Tassoul \& Tassoul (\cite{tassoul}) on the other hand appears to 
be much too efficient. Nevertheless, care has to be taken interpreting these 
results, because both theories give timescales for the synchronisation of entire
 stars from the core to the surface, while only the rotation at the surface can 
be measured from line broadening. Goldreich \& Nicolson 
 (\cite{goldreich}) showed that in stars with radiative envelopes and Zahn's 
 braking mechanism at work, the synchronous rotation proceeds from the surface 
 towards the core of the star. This means that the outer layers are 
 synchronised faster than the rest of the star. This effect would explain the 
 discrepancy between Zahn's theory and our results at least to a certain 
 extent. Unfortunately it was not possible to quantify this effect so far 
 (see e.g. review by Zahn \cite{zahn3}). 

Tidal synchronisation does not necessarily lead to an equality of orbital and 
rotational period. Higher spin resonances are possible and would change the 
derived parameters significantly (in case of the planet Mercury the ratio of 
orbital and rotational period is $3/2$). To fall into a higher resonance, the 
binary eccentricity has to be high at some point of its evolution. But close 
sdB binaries underwent at least one common envelope phase (maybe two in case 
of compact companions), which led to a circularisation of the orbit. The small 
eccentricities in some of our programme binaries reported by Edelmann 
et al. (\cite{edelmann}) and Napiwotzki et al. (in prep.) are considered to be still consistent with this 
scenario. For these reasons, higher resonances are unlikely to occur in this 
evolutionary channel.

\section{Empirical evidence for synchronisation}\label{sec:empirical}

The timescale of the synchronisation process is highly dependent on the tidal 
force exerted by the companion. If the companion is very close and the orbital 
period therefore very short, synchronisation is established much faster than 
in binaries with longer orbital periods. If an sdB binary with given orbital 
period is proven to be synchronised, all other sdB binaries with shorter 
orbital periods should be synchronised as well. Although the timescales also 
scale with sdB radius and companion mass, the orbital period is the dominating
factor because sdB radii differ only little and the dependence on 
companion mass is not so strong.

\subsection{Eclipsing and ellipsoidal variable systems}

Eclipsing sdB binaries are of utmost importance to test the synchronisation 
hypothesis because the inclinations can be derived directly from their 
light curves. It has been shown in Sect.~\ref{sec:lowmassm} that the parameters of the 
eclipsing sdB+dM binaries PG\,1336$-$018, HS\,0705$+$6700 and HW\,Vir are consistent 
with synchronised orbits. This essentially means that the calculated 
$v_{\rm rot}\sin{i}$ for synchronous rotation, which can be obtained as 
described in Sect.~\ref{sec:ana} given the orbital period, the radius of the 
sdB and the inclination angle are known, is consistent with the measured value. 
In eclipsing systems, all these parameters can be measured.

This provides clear empirical evidence that at least the upper layers of the 
stellar envelopes are synchronised to the orbital motion of the eclipsing sdB 
binaries in our sample. We therefore conclude that all sdBs in close binaries 
with orbital periods up to $0.12\,{\rm d}$ should be synchronised as well.

Two well studied sdBs clearly show ellipsoidal variations in their light curves with
 periods exactly half the orbital periods (KPD\,1930+2752, Bill\`{e}res et al. \cite{billeres00}, 
Maxted et al. \cite{maxted2}, Geier et al., 
 \cite{geier}, is further discussed in Sect.~\ref{sec:lowmasswd}; 
 KPD\,0422+5421, Koen et al. \cite{koen4}, Orosz \& Wade \cite{orosz}, is not part of our sample). 
 This alone is only an indication for tidal synchronisation, because the 
 light curve variations have to be present at the proper orbital phases as 
 well. To really prove synchronisation it is necessary that the stellar 
 parameters determined independently from the light curve analysis are 
 consistent with a synchronised orbit. This is the case for KPD\,0422$+$5421 as 
 well as KPD\,1930$+$2752. Both ellipsoidal variable systems have very short 
 periods of $\simeq0.1\,{\rm d}$ and high inclination. Otherwise ellipsoidal 
 variations are very hard to detect. 

Most compelling evidence for synchronisation in a binary system with a period considerably longer than that of the above mentioned systems is provided in the case of the eclipsing sdB+WD binary KPD\,1946$+$4340 ($P=0.404\,{\rm d}$). 
Bloemen et al. (\cite{bloemen}) derived most accurate binary parameters from a spectacular high-S/N light 
curve obtained by the Kepler mission. These results are fully consistent with the constraints we put on this 
system (see Sect.~\ref{sec:lowmasswd}). We therefore conclude that sdB binaries with periods shorter than $P\simeq0.4\,{\rm d}$ should be synchronised.

Furthermore, the sdB+WD binary PG~0101$+$039 ($P=0.567\,{\rm d}$) shows very weak luminosity
 variations at half the 
orbital period detected in a 16.9 day long, almost uninterrupted light curve 
obtained with the MOST satellite (Randall et al. \cite{randall2}). Geier 
et al. (\cite{geier2}) showed that the sdB in this binary is most likely
 synchronised. The empirical lower limit for tidal synchronisation in close 
 sdB binaries is therefore raised to $P\simeq0.6\,{\rm d}$.

\subsection{Asteroseismology}

An independent method to proof orbital synchronisation is provided by 
asteroseismology. Van Grootel et al. (\cite{vangrootel}) were able to 
reproduce the main pulsation modes of the short period pulsating sdB in the binary 
Feige 48 ($P\simeq0.38\,{\rm d}$), derived the surface rotation from the 
splitting of the modes and concluded that the subdwarf rotates synchronously.

Charpinet et al. (\cite{charpinet5}) reach a similar conclusion for the short 
period eclipsing binary PG\,1336$-$018 ($P\simeq0.10\,{\rm d}$). Furthermore 
they probed the internal rotation of the star below the surface layers by 
applying a  differential rotation law and showed that the sdB rotates as a 
rigid body at least down to $0.55\,R_{\rm sdB}$. The remarkable consistency 
of the binary parameters derived by asteroseismology (Charpinet et al. 
\cite{charpinet5}), binary light curve synthesis (Vu\v ckovi\'c et al. 
\cite{vuckovic2}) and the analysis presented here has to be pointed out 
again (see Sect.~\ref{sec:lowmassm}). Asteroseismic ana\-lyses revealed that 
sdB binaries up to orbital periods of about $0.4\,{\rm d}$ are synchronised. 
We therefore conclude that all sdBs in close binaries with shorter periods 
should be synchronised as well.

\section{Synchronisation challenged}\label{sec:challenge}

In Sect.~\ref{sec:syncsample} we have shown that synchronisation in our sample has been
established for binaries with periods below $\simeq1.2\,{\rm d}$. 
This is corrobated by the theory of synchronisation 
although different version of the theory give vastly different results. Empirical evidence sets a
limiting period of $0.6\,{\rm d}$. About half of our sample has periods below that limit
and should therefore be synchronised. These arguments are correct for the sample 
but may not hold for individual objects. 
We envisage two options: The subdwarf may not be core helium-burning (Sect.~\ref{sec:hd188}). Or an individual EHB star may be too young to have reached synchronisation (Sect.~\ref{sec:age}). 

\begin{figure}[t!]
\begin{center}
	\resizebox{\hsize}{!}{\includegraphics{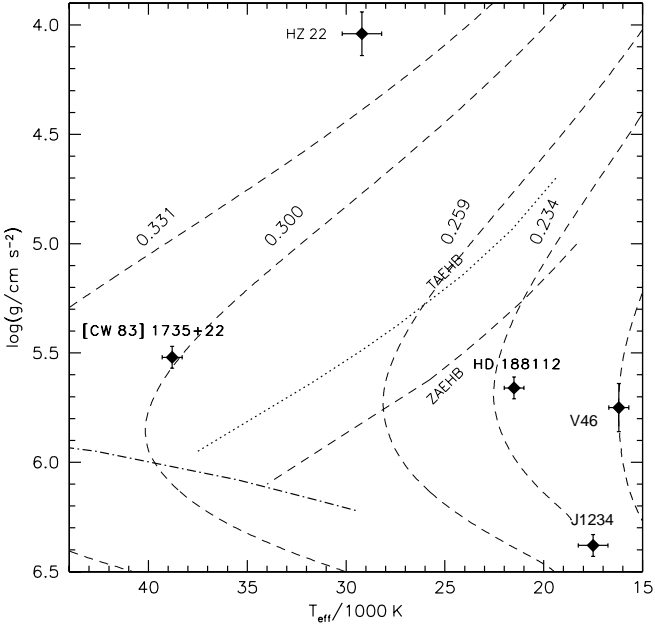}}
	\caption{$T_{\rm eff}-\log{g}$-diagram similar to Fig.~\ref{fig:tefflogg}. The black filled diamonds mark the known post-RGB binaries HD\,188112 (Heber et al. \cite{heber5}), NGC\,6121$-$V46 (V46 for short, O'Toole et al. 
	 \cite{otoole}), HZ~22 (Sch\"onberner, \cite{schoenberner}, Saffer et
	 al., \cite{saffer97}) 
	 and SDSS\,J123410.37$-$022802.9 (J1234 for short, 
	 Liebert et al. \cite{liebert}). The candidate post-RGB system [CW83]~1735$+$22 is included as well. 
	 The helium main sequence and the
	EHB-band 
	are superimposed with post-RGB evolutionary tracks from Driebe et al.
	 (\cite{driebe}) labelled by their masses.
	 }
	\label{teffloggpostRGB}
\end{center}
\end{figure}

\begin{figure}[t!]
\begin{center}
	\resizebox{\hsize}{!}{\includegraphics{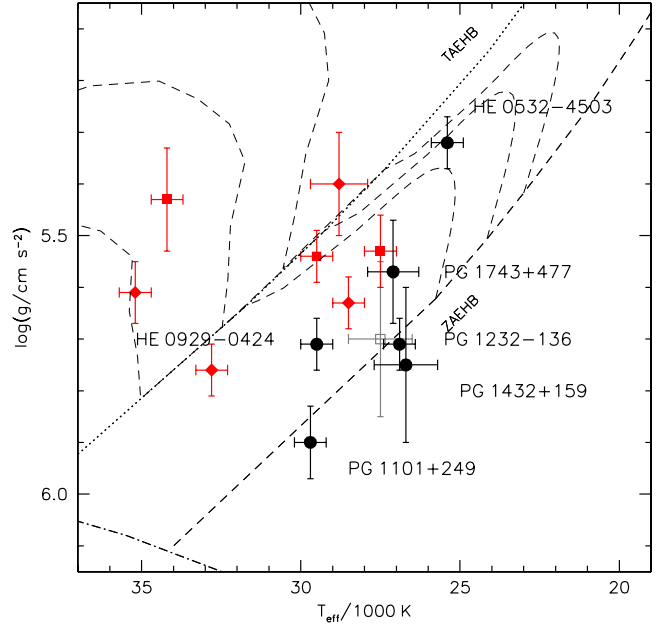}}
	\caption{$T_{\rm eff}-\log{g}$-diagram, same as 
	Fig.~\ref{fig:tefflogg} but restricted to the massive systems (black filled circles) described in 
	section~\ref{sec:highmassbh} and supplemented 
	by short-period systems ($\simeq0.1\,{\rm d}$, filled diamonds) 
	where synchronisation 
	has been proven empirically. The two filled squares mark the longer 
	period binaries Feige\,48 ($\simeq0.38\,{\rm d}$), KPD\,1946$+$4340 ($\simeq0.40\,{\rm d}$) and PG\,0101$+$039 
	($\simeq0.57\,{\rm d}$), which are known to be synchronised. The 
	open square marks the non-synchronised binary PG\,2345$+$318. 
	The helium main sequence and the EHB band 
	are superimposed with EHB evolutionary tracks from Dorman et al. 
	(\cite{dorman}) labelled by their masses.  
	}
	\label{teffloggmassive}
\end{center}
\end{figure}

\subsection{[CW\,83]\,1735$+$22 and post-RGB evolution}\label{sec:hd188}

The only sdB star known not to burn helium in the core is the single-lined close binary HD\,188112 (Heber et al. \cite{heber5}). According to its atmospheric parameters it is situated well below the EHB (see Fig.~\ref{teffloggpostRGB}). By interpolation of evolutionary tracks from Driebe et al. (\cite{driebe}) a mass of $0.23\,M_{\rm \odot}$ was derived, which could be verified directly, because an accurate parallax of this object was obtained by the Hipparcos satellite. 

The different evolution of 
so called post-RGB objects like HD\,188112 compared to EHB stars should affect their rotational properties. Post-RGB stars constantly shrink during their evolution towards the WD cooling tracks. Since these stars are not expected to lose angular momentum during the contraction, they have to spin up. In contrast to this a core helium-burning sdB star expands by a factor of about two within $\simeq100\,{\rm Myr}$ and is expected to spin down. Besides HD\,188112 some other objects are also considered to belong to this class (see Fig.~\ref{teffloggpostRGB}).

The post-RGB scenario may explain the unusual properties, especially the fast rotation, of the sdB binary {\bf [CW\,83]\,1735$+$22} (see Sect.~\ref{sec:syncsample}). The star is among the hottest in our sample and it lies far from the EHB band (see Fig.~\ref{teffloggpostRGB}). According to the mass tracks of Driebe et al. (\cite{driebe}) [CW\,83]\,1735$+$22 would have a mass of about $0.3\,M_{\rm \odot}$ (see Fig.~\ref{teffloggpostRGB}). Such a star should shrink by a factor of $5.5$ within $0.3\,{\rm Myr}$ (Driebe et al. \cite{driebe}), which is much shorter than the synchronisation time. Hence we regard its high projected velocity as strong evidence that [CW\,83]\,1735$+$22 is a post-RGB star just like HD\,188112. Since the lifetime of such an object is predicted to be only a few million years, such stars should be rare. The predicted low mass of [CW\,83]\,1735$+$22 can be verified in the way described in Heber et al. (\cite{heber5}) as soon as the GAIA mission will have measured an accurate trigonometric parallax of this star. 

One may speculate that the different rotational properties of post-RGB stars may have an influence on the synchronisation process if  they are in close binary systems. The spin-up caused by the shrinkage of the star may counteract the spin-down caused by the tidal influence of the companion. Should post-RGB stars have longer synchronisation timescales than EHB stars this may be invoked as a 
convenient explanation for the putative high fraction of sdB binaries with massive compact companions. If these binaries should have post-RGB primaries and should not be synchronised, the derived companion masses would be wrong. 

This scenario is considered to be unlikely. First of all, we would expect post-RGB stars to rotate faster than synchronised. For the putative sdB+NS/BH systems low projected rotational velocities are measured. If the sdBs should rotate even faster than synchronised, the inclination angle would be even lower and the derived companion masses would go up. 

Another strong argument against a post-RGB nature of the sdBs in the candidate systems with massive compact companions is their location in the $T_{\rm eff}-\log{g}$ diagram (see Fig.~\ref{teffloggmassive}). All these binaries are found on or near the EHB, while the known post-RGB stars are obviously not concentrated near the EHB (see Fig.~\ref{teffloggpostRGB}). We therefore conclude that the sdBs with putative massive compact companions are post-EHB rather than post-RGB stars.

\subsection{The role of the stellar age}\label{sec:age}

Up to now we have assumed that the sdB stars already have spent a significant 
part of their total life time on the EHB. 
In the canonical picture it might be possible to estimate the age of an
individual star by comparing its position in the (T$_{\rm eff}$, $\log
g$)-diagram to EHB evolutionary tracks (e.g. Dorman et al. \cite{dorman}), as
the core mass is fixed at the core helium flash. In binary population models,
however, a degeneracy between mass and age arises as there is a spread of sdB 
masses (see Zhang et al. \cite{zhang}). 

Because our sample stars nicely populate the canonical EHB band (see 
Figs.~\ref{fig:tefflogg}, \ref{fig:teffloggsync}, \ref{fig:teffloggnosync}), 
we shall assume that a star is young if it is on or 
close to the zero-age extreme horizontal branch (ZAEHB) and old if not. 
Note that the speed of evolution along the EHB tracks is nearly constant.

We shall now explore whether some of our targets might possibly be too young
to be synchronised. We shall start with PG\,2345$+$318 and inspect the sample in
the light of the lesson to be learnt.

\subsection{PG\,2345$+$318}

{\bf PG\,2345$+$318} is a short period ($0.24\,{\rm d}$) sdB binary. 
We derive a high companion mass of $1.9\pm0.7\,M_{\rm \odot}$ at an 
inclination angle of about $22^{\circ}$ indicating that the companion is another massive compact object, i.e. a neutron star or a massive white dwarf. At such a low inclination eclipses are not expected to occur. 
 
However, Green et al. (\cite{green}) presented a preliminary light curve of 
this star, and detected a shallow eclipse probably by a white dwarf.\footnote{Besides KPD\,0422$+$5421 (Orosz \& Wade 
\cite{orosz}), PG\,0941$+$280 (Green et al. \cite{green}) and KPD\,1946$+$4340 (Bloemen et al. \cite{bloemen}) this is just the 
fourth such system known.} Without the additional information from the light 
curve this object would therefore be identified as another candidate sdB binary with massive 
  compact companion. The detection of eclipses immediately rules out this 
  scenario. The inclination angle has to be near $90^{\circ}$ and the 
  companion a white dwarf with a mass of $0.38\,M_{\rm \odot}$ according to the constraint set 
   by the binary mass function. This means that the sdB star in this binary 
   rotates more slowly than synchronised and proves that such objects exist 
among binaries with short orbital periods. The most reasonable explanation for this may be that the 
system is very young and the synchronisation process not finished yet.

The atmospheric parameters of this star (see Table~\ref{orbit}) place it indeed near the zero-age EHB (Fig.~\ref{teffloggmassive}), although they have somewhat larger errors than most other stars due to the lack of high quality low resolution spectra (Saffer et al. \cite{saffer}). But the light curve presented by Green et al. (\cite{green}) reveals more 
information, which corrobate this scenario. An interesting feature is the presence of a
 shallow reflection effect and a weak secondary minimum, which provides 
 evidence that the white dwarf contributes significantly to the optical flux. This in 
 turn means that the white dwarf must be young (assuming a luminosity of
  $0.5\,L_{\rm \odot}$ evolutionary tracks imply an age of the order of
   $10^{6}\,{\rm yr}$) and is another piece of evidence that the system is too young to be 
   synchronised. Since no light curve solution for PG\,2345$+$318 is published 
   yet, the discussion of this object must remain preliminary. 

\subsection{Are the systems with massive compact companions too young to be
synchronised?}\label{sec:young}

What are the implications for our candidate sample of sdB binaries with 
massive compact companions? The orbital periods of these binaries range from 
$0.26\,{\rm d}$ to $0.52\,{\rm d}$ where synchronisation should be established 
according to the results presented in Sect.~\ref{sec:syncsample} and \ref{sec:empirical}. Given these short orbital periods, the
 binaries in question should be synchronised.  

Even if the candidate systems were bona-fide EHB stars, 
they may just be too young to be synchronised. In Fig.~\ref{teffloggmassive} we plot the positions of the candidate systems with compact companions and compare them to the calibrators Feige\,48, PG\,0101$+$039 and KPD\,1946$+$4340.
It is obvious that the first two of these synchronised sdBs lie closer to the terminal age EHB than to the
zero age EHB. KPD\,1946$+$4340 is already evolved from the EHB and most likely burning helium in a shell. These are indications that these binaries are relatively old. We also note that the 
position of PG\,1743$+$477 nearly coincides with that of PG\,0101$+$039. From this coincidence we would expect
it to be synchronised and, hence, the constraint on the companion mass to be reliable.

We also plot the position of the non-synchronised system 
PG\,2345$+$318 in Fig.~\ref{teffloggmassive} which lies near the zero-age EHB. 
PG\,1232$-$136 and PG\,1432$+$159 are found close to PG\,2345$+$318 and near
the zero-age EHB and thus may be rather young as well. The same holds for  
PG\,1101$+$249 which is considerably hotter but also situated very near 
the zero-age EHB (ZAEHB). 

The remaining candidate sdB binaries with putative massive compact companions are in a similar evolutionary stage as the 
synchronised systems in the middle of the EHB band. We conclude that some but not all sdBs in the candidate systems could be too young to have reached synchronisation.

\section{Summary and Outlook}\label{sec:summary}

We have analysed a sample of 51 sdB stars in close single-lined binary systems.
This included 40 systems for which the orbital parameters have been determined 
previously. The subsample comprises half of all systems known so far. 
From high resolution spectra taken with different instruments the projected rotational velocities of these stars have been derived to an unprecedented precision. Accurate measurements of the surface gravities have mostly been 
taken from literature. 
 Assuming orbital synchronisation and an sdB mass distribution as suggested by
 binary population synthesis models as well as by asteroseismology, 
 the masses and the nature of the unseen 
 companions could be constrained in 31 cases. 
 Only in five cases we were unable to classify unambiguously. These companions
 may either be low mass main-sequence stars or white dwarfs. 
 The companions to seven sdBs could be clearly 
 identified as late M stars. One binary may have a brown dwarf companion. 
 The unseen companions of nine sdBs are white dwarfs with typical masses, one WD companion has a very low mass.
   
In eight cases (including the well known system KPD1930$+$2752) 
the companion mass exceeds $0.9\,M_{\rm \odot}$. Four of the companions even
exceed the Chandrasekhar limit indicating that they may be neutron 
stars; even a stellar mass black hole is possible for the most massive
companions. 

The basic assumption of orbital synchronisation in close sdB binaries has been 
discussed in detail. Our analysis method yielded consistent 
 results for binaries up to an orbital period of $\simeq1.2\,{\rm d}$.
Theoretical timescales for synchronisation were calculated using two different 
approaches. The theory of Zahn (\cite{zahn2}) was found to be too inefficient
while that of Tassoul \& Tassoul (\cite{tassoul}) predicts too short timescales. 
The predictions from both theories are strongly discrepant, calling for 
empirical constraints.
 
Independent observational evidence for synchronisation in sdB binaries 
comes from light curve analyses of eclipsing, ellipsoidal deformed, and pulsating sdBs. 
Due to this evidence sdB binaries with periods shorter than 
$\simeq0.6\,{\rm d}$ should be synchronised. This includes all of the putative
massive systems.

Hence, an evolutionary model for the origin of sdB stars with neutron star or
black hole companions was devised indicating that common envelope evolution is
indeed capable of producing such systems, though at a lower rate than observed.
An appropriate formation channel includes two phases of unstable 
 mass transfer and one supernova explosion.

The distribution of the inclinations of the systems of normal mass appears to 
be consistent with expectations, whereas a lack of high inclinations became 
obvious for the massive systems. 

There is one star in the sample which rotates fast despite its rather long orbital period. This as well as its 
position far from the EHB band hints at a post-RGB nature. The post-RGB stars are expected to be spun-up due to their ongoing 
contraction.

The larger number of putative massive companions in low inclination systems is puzzling.
Therefore, we investigated alternative interpretations. The fraction of massive unseen companions can only be lowered, if the sdBs themselves have masses much lower than the anticipated range of 
$0.43 - 0.47\,M_{\rm \odot}$ for EHB stars. Evolutionary calculations 
showed that EHB stars with masses as low as 
$0.30\,M_{\rm \odot}$ can be formed if helium ignites under non-degenerate 
conditions but should be very rare. Assuming such low sdB masses, only one unseen companion remains 
more massive than the Chandrasekhar limit.  
This fraction of $3\%$ is roughly 
consistent with theoretical predictions. 
Whether the sdB mass is small or not can be checked directly as 
soon as accurate parallaxes of 
these relatively bright stars will become available through the GAIA mission.

The putative massive sdB systems might not be synchronised if their age
is much less than anticipated. That this can happen is witnessed by 
PG\,2345$+$318, a short-period sdB binary in our sample, that we would have
classified as a low-inclination massive system as well, if it were not proven
by eclipses to be highly inclined. Hence the system is not synchronised 
despite of its short period ($0.24\,{\rm d}$).
Due to a degeneracy between mass and age, it is 
difficult to estimate the sdB's age without knowing its mass. Adopting the
canonical mass, we nevertheless estimated the stars' ages from their position 
in the EHB band. Indeed, PG\,2345$+$318, is located right on the zero-age EHB
as are the massive candidates PG\,1232$-$136, PG\,1432$+$159 and PG\,1101$+$249.
These stars may possibly be too young to have reached synchronisation. Hence
the companion masses we derived would be spurious. 
However, there is no indication that the other massive systems could be young.  

Even if we dismiss three candidates because they may be too young and assume
that the others are of low mass, PG\,1743$+$477 and, in particular,
HE\,0532$-$4503 remain as massive candidates whose companions have masses close
to or above the Chandrasekhar mass. 

Different approaches may be chosen to directly verify the presence of neutron 
star or black hole companions in our candidate systems. 
None of the sdBs in our target systems fills its Roche lobe. No mass transfer 
by Roche lobe overflow to the unseen companion can occur and
therefore no X-ray emission is expected. 
The ROSAT all-sky survey catalogue (RASS, Voges et al. \cite{voges}) has been checked and, indeed, no sources have been detected at the positions of any candidate sdB+NS/BH systems. 
The detection limit of this survey reaches down to about $10^{-13}\,{\rm erg\,cm^{-2}s^{-1}}$. 
However, sdB stars are expected to have weak winds. Hence accretion from the
sdB wind might result in faint X-ray emission. This occurs in the bright sdO+WD system HD\,49798 (Mereghetti et al., \cite{mereghetti}). Although stellar wind mass loss rates in sdBs are predicted to be small ($<10^{-12}\,M_{\rm \odot}{\rm yr^{-1}}$, e.g. Vink \& Cassisi \cite{vink}; Unglaub \cite{unglaub3}), they may be sufficient 
to cause detectable X-ray flux powered by wind accretion.  X-ray telescopes like Chandra or XMM-Newton may 
be sensitive enough to detect such weak sources. Pulsar signatures of rapidly spinning 
 neutron star companions may be detectable with radio telescopes.

Tidal forces by the companion cause an ellipsoidal deformation of the primary 
in close binary systems. This deformation appears as a variation of light at 
half the orbital period. Two very close subdwarf binaries with orbital periods 
of $\simeq\,2\,{\rm h}$ and high orbital inclination show light variations of 
about $1\%$, which can be detected from the ground. Performing binary 
light curve synthesis it was possible to derive the masses of the binary 
components (Orosz \& Wade \cite{orosz}; Geier et al. \cite{geier}). Signatures 
of ellipsoidal deformation in the light curves of binaries with longer orbital 
periods and lower inclination are much weaker ($\simeq0.01\%$, Drechsel priv. 
comm.; Napiwotzki et al. in prep.) and therefore not detectable from the ground. The existence of such very
 shallow variations has been proven for the subdwarf binary PG\,0101$+$039 with
  an orbital period of $13.7\,{\rm h}$ using a light curve of almost 
  $17\,{\rm d}$ days duration taken with the MOST satellite. The ellipsoidal 
  variation was found to be $0.025\%$ (Geier et al. \cite{geier2}). 
The full potential of high precision photometry for the analysis of sdB binaries has most 
recently been demonstrated by Bloemen et al. (\cite{bloemen}), who analysed a Kepler light 
curve of the eclipsing sdB+WD binary KPD\,1946$+$4340. High precision light curves of the best candidates in our sample should be measured with HST. The nature of their unseen companion could then be clarified.

Most of the candidate massive systems have low orbital inclination. 
High inclination systems must exist as well. In this case a determination of the orbital parameters is 
sufficient to put a lower limit to the companion mass by calculating the 
binary mass function. If this lower limit exceeds the Chandrasekhar mass and no sign of a companion 
is visible in the spectra, the existence of a massive compact companion is 
proven without making any additional assumptions. The Hyper-MUCHFUSS project 
(Hypervelocity stars or Massive Unseen Companions to Hot Faint Underluminous 
Stars from SDSS, Geier et al. in prep.) was launched in 2007. One of the aims 
of this project is to search for sdB binaries with massive compact companions 
at high inclinations in a sample of stars selected from the SDSS data base. 

\begin{acknowledgements}

We would like to thank Z. Han for providing us with stellar structure models 
of sdB stars. We thank E. M. Green, N. Reid and L. Morales-Rueda for sharing 
their data with us. We are grateful to R. H. \O stensen and S. Bloemen, who provided us with 
informations about new detections or non-detections of indicative features in 
sdB light curves, as well as H. Drechsel for modelling such light curves for us. 
S. G. was supported by the Deutsche Forschungsgemeinschaft under grant 
He~1354/40-3. Travel to La Palma for the observing run at the WHT was funded by DFG 
through grant He 1356/53-1.

\end{acknowledgements}

\end{document}